\documentclass[%
 aip,
 jmp,
amsmath,amssymb,
reprint,%
]{revtex4-1}

\usepackage{graphicx}

\usepackage{color} 
\usepackage{bm}
\usepackage[utf8]{inputenc}
\usepackage[T1]{fontenc}
\usepackage{mathptmx}
\usepackage{soul} 

\usepackage{cancel}

\usepackage{hyperref}

\newcommand{\ie}{\textit{i.e.}}

\newcommand{\rmd}{\mathrm{d}}
\newcommand{\rmi}{\mathrm{i}}
\newcommand{\rme}{\mathrm{e}}
\newcommand{\tr}{\mathrm{tr}}

\newcommand{\la}{\langle}
\newcommand{\ra}{\rangle}

\newcommand{\One}{\openone}
\newcommand{\eps}{\varepsilon}
\newcommand{\h}{\hat}
\newcommand{\rmw}{\mathrm{w}}

\newcommand{\vk}{\varkappa}

\newcommand{\bea}{\begin{eqnarray}}
\newcommand{\eea}{\end{eqnarray}}

\newcommand{\be}{\begin{equation}}
\newcommand{\ee}{\end{equation}}

\newcommand{\AY}[1]{{#1}}

\begin{document}


\title[Generalized symmetries, invariant solutions and conservation laws in the Jaynes-Cummings model]
{Generalized symmetries, invariant solutions and conservation laws in the Jaynes-Cummings model}

\author{L. M. Piñuelas Castro}
\affiliation{Departamento de Ciencias Naturales y Exactas, CUValles, Universidad de
 Guadalajara, Carretera Guadalajara - Ameca Km. 45.5 C.P. 46600, Ameca,
 Jalisco, M\'exico.}

\author{P. C. L\' opez V\' azquez}
\affiliation{Departamento de Ciencias Naturales y Exactas, CUValles, Universidad de
 Guadalajara, Carretera Guadalajara - Ameca Km. 45.5 C.P. 46600, Ameca,
 Jalisco, M\'exico.}

\author{A. Yakhno}
\altaffiliation{Corresponding Author}
\email{alexander.yakhno@academicos.udg.mx}
\affiliation{Departamento de Matem\'aticas, CUCEI, Universidad de Guadalajara,
 Blvd. Marcelino Garc\'\i a Barragan y Calzada Ol\'\i mpica,
 C.P. 44840, Guadalajara, Jalisco, M\' exico.}%

\date{\today}

\begin{abstract}
In this study, we investigate the Jaynes--Cummings model (JCM) using Lie symmetry analysis and 
conservation-law theory. The dynamics is formulated as a system of partial differential equations 
by projecting the von Neumann equation onto the atomic degrees of freedom and representing the field 
mode through its characteristic function. We determined the admitted point and generalized symmetries 
and constructed invariant solutions 
\AY{by imposing quantum-mechanical constraints}.
The conventional dressed-state dynamics is recovered, while a second class of 
solutions with radial dependence expressed through Heun polynomials is obtained for coupled 
atom--field configurations.

We also applied a generating-function methodology to derive the local conservation 
laws of the JCM differential system \AY{in Fourier phase space}. In addition to recovering 
\AY{the standard 
excitation-number constant of motion}, we obtained additional conserved currents involving atomic 
populations, coherence, reduced-state purity, \AY{and derivatives of the field characteristic functions}. 
\AY{One of these currents yields, at the Fourier-space origin, a balance equation for a combination 
of atomic purity and coherence whose evolution is driven by atom--field coupling and joint atom--field moments. 
For globally pure states, the purity contribution provides an indirect connection with the development 
of atom--field entanglement.} 
The symmetry structure generates generalized symmetries and an infinite hierarchy of 
conservation laws.
\end{abstract}

\maketitle

\section{\label{intro}Introduction}
The Jaynes--Cummings model (JCM), first introduced in 1963 by Edwin Jaynes
and Fred Cummings~\cite{JC63}, is one of a \AY{models} of quantum
light--matter interaction. It describes the coherent exchange of excitations
between a two-level atom and a single quantized field mode and has played a
central role in quantum optics, cavity quantum electrodynamics, and quantum
information science~\cite{Ge04,QoptFM06,NC10,Purcell1995,Haroche06,Wallraff2004,Duan04,Fink2008}.
Its interaction has no direct classical counterpart in the sense that energy
exchange occurs in discrete quanta, whereas the eigenstates of the coupled
Hamiltonian form the well-known dressed-state basis, which is generally
entangled across the atom and field degrees of freedom~\cite{Frasca99}. These
features make the JCM a particularly suitable setting in which to investigate
how mathematical structures, such as symmetries and conservation laws, reflect
the underlying quantum dynamics.

The conventional analytical treatment of the JCM is based on the decomposition
of the dynamics into invariant excitation subspaces and the diagonalization
of the Hamiltonian within each of them. Although this construction provides a 
standard dressed-state solution, the dynamics can also be formulated as a
coupled system of partial differential equations (PDEs) for the atomic components and field mode in the characteristic-function representation~\cite{Ozo04,Lopez20}. This formulation retains the full atom--field structure
while providing a natural framework for studying the model from the perspective
of differential equations.

In this study, we explore what Lie symmetry theory reveals about this PDE
representation of the JCM. Symmetry analysis identifies the continuous
transformation groups admitted by the dynamical equations and uses them to
construct invariant solutions, reduce the number of independent variables,
and expose structural relations that are not evident from the Hamiltonian
diagonalization alone~\cite{Hydon_2000,Ovsiannikov82,Olver1986}. In this sense,
the objective is not to replace the standard solution of the JCM but to
characterize it from a symmetry-based perspective and to determine whether the
admitted symmetries produce additional classes of physically admissible
solutions.

The group-analysis procedure is naturally semi-inverse. One first derives the
solution forms imposed by a selected symmetry and subsequently determines the
initial and boundary conditions compatible with them. In the present problem,
the physical admissibility is fixed by the quantum-mechanical conditions imposed on
the characteristic function and reduced atomic density matrix. This
distinction is essential: differential equations may admit mathematically
valid invariant solutions that do not represent physical quantum states. Therefore, symmetry analysis provides both a classification of invariant
solutions and a criterion for determining which of them survive the physical
constraints of the model.

The application of this method recovers familiar dressed-state dynamics as
one of the invariant solutions of the PDE system. At the same time, another
subalgebra of group generators leads to a distinct class of invariant solutions whose radial
dependence is expressed in terms of Heun polynomials. These solutions
extend beyond the conventional dressed-state construction and are compatible
with generic coupled atom--field configurations, although their complete
physical interpretation remains to be established. Their appearance illustrates
one of the main outcomes of the symmetry approach: It can reveal admissible
analytical structures that are not naturally suggested by the usual
Hamiltonian treatment.

The differential formulation also provides a natural setting for the analysis
of nontrivial conservation laws. For this purpose, we employed the methodology
developed by Vinogradov's school, which relates the conserved currents of a system of
differential equations to their generating functions
~\cite{Vinogradov:1984aa,Vinogradov1989_1,Vinogradov1999}. Within this
framework, the standard conservation of the total number of excitations is  
recovered directly from the PDE system. In addition, further conserved currents
were obtained, whose densities and fluxes involve atomic populations, coherence,
purity, and characteristic-function moments associated with field
dynamics.

Of particular interest is the conservation-law balance equation for the quantity
$P_a-f_a^2$, where $P_a$ is the purity of the reduced atomic state, and $f_a$
is its coherence function. Its time variation is determined by the atom--field
coupling strength and by combinations of atomic expectation values with
derivatives of the field characteristic functions. This result does not imply
that entanglement itself is directly conserved. Rather, it identifies a
conserved current involving quantities whose evolution is linked to the
redistribution of purity, coherence, and correlations between the atom and the
field. For globally pure atom--field states, changes in local purity are directly
associated with the development of bipartite entanglement, which gives these
conservation laws a natural connection to the entanglement dynamics of the
JCM. A related association between conserved quantities and entanglement
dynamics was also suggested in Ref.~\onlinecite{Lopez26}.

Therefore, this study aims to examine the mathematical and physical
outcomes of applying symmetry analysis and conservation-law theory to the JCM.
The method provides a different characterization of known solutions, reveals an
additional class of invariant solutions, recovers the standard excitation-number
constant of motion, and produces non-trivial conserved currents associated with
fundamental quantum properties of the atom--field dynamics. To the best of our
knowledge, this is the first application of Vinogradov's conservation-law
framework to a composite quantum system, suggesting a possible route for the
analysis of more complex quantum models formulated as systems of differential
equations.

This paper is organized as follows. In Sec.~\ref{JCeqs}, we derive the
dynamical equations of the JCM in the characteristic-function representation
and introduce the transformations used throughout the analysis. A brief review
of the conventional Hamiltonian and dressed-state construction is presented in
Appendix~\ref{appJCM}. In Sec.~\ref{LST}, we determine the admitted point and
generalized symmetries and derive the invariant solutions associated with the
relevant subalgebras. In Sec.~\ref{CL}, we construct the corresponding
conservation laws, including the excitation-number conservation law and the
currents related to atomic purity, coherence, and atom--field correlations.
Finally, in Sec.~\ref{summ}, we summarize the main results and their possible
implications.

\section{\label{JCeqs}Differential equations of the JCM}
In this section, we derive a system of differential equations that governs the
dynamics of the JCM in the characteristic function representation. We begin with 
the von Neumann equation
\begin{equation}\label{vNeu}
  \rmi {\rmd \over \rmd t}\varrho = [\h{H},\varrho]\,,
\end{equation}
where $\varrho$ is the density operator of the composite atom-field system and
$\h{H}$ is the JCM Hamiltonian given by: 
\begin{equation}\label{ham}
  \h{H} =
  \frac{\Delta}{2}\sigma_z\otimes \One
  + \One\otimes\h{n}
  + g\left( \sigma_{+}\otimes\h{a}
  + \sigma_{-}\otimes\h{a}^{\dag} \right).
\end{equation}
Here, $g=\lambda/(2\omega_o)$ is the dimensionless atom-field coupling
strength, \AY{$\Delta=\omega_q/\omega_o$, $\omega_q$ is the atomic level spacing measured in units of frequency,
$\omega_o$ is the characteristic frequency of the field mode, also described in the same units,}  $\sigma_+$ and $\sigma_-$ are the atomic raising and lowering
operators, and $\h{a}^{\dag}$ and $\h{a}$ are the creation and annihilation
operators of the field mode, respectively. The interaction term describes the coherent
exchange of one excitation between the atom and the field: the absorption of a
field quantum raises the atom, whereas emission by the atom increases the field
excitation number.
This formulation allows
one to describe both pure and mixed quantum states, and is therefore more general
than a state-vector description. Moreover, rather than working on the eigenbasis of the total Hamiltonian, we project
Eq.~(\ref{vNeu}) onto the bare atomic basis,
$\{|e\rangle\otimes\One_f, |g\rangle\otimes\One_f\}$. In this basis,
\[
\sigma_z = |e\rangle\langle e|-|g\rangle\langle g|,\quad
\sigma_+ = |e\rangle\langle g|,\quad
\sigma_- = |g\rangle\langle e|=\sigma_+^\dagger .
\]

Writing the density operator in atomic blocks,
$\varrho_{ij}=\langle i|\varrho|j\rangle$, with
$i,j=e,g$, yields the following system of operator equations:
\begin{eqnarray}
  \rmi \dot{\varrho}_{ee}
  &=&
  [\h{n}+1/2,\varrho_{ee}]
  +g\left(\h{a}\varrho_{ge}-\varrho_{eg}\h{a}^{\dag}\right),
  \nonumber\\[0.5em]
  \rmi \dot{\varrho}_{eg}
  &=&
  \Delta\varrho_{eg}
  +[\h{n}+1/2,\varrho_{eg}]
  +g\left(\h{a}\varrho_{gg}-\varrho_{ee}\h{a}\right),
  \nonumber\\[0.5em]
  \rmi \dot{\varrho}_{ge}
  &=&
  -\Delta\varrho_{ge}
  +[\h{n}+1/2,\varrho_{ge}]
  +g\left(\h{a}^{\dag}\varrho_{ee}-\varrho_{gg}\h{a}^{\dag}\right),
  \nonumber\\[0.5em]
  \rmi \dot{\varrho}_{gg}
  &=&
  [\h{n}+1/2,\varrho_{gg}]
  +g\left(\h{a}^{\dag}\varrho_{eg}-\varrho_{ge}\h{a}\right).
  \label{meq4}
\end{eqnarray}

To apply the Lie symmetry theory, we move from this operator formulation to the
characteristic-function representation of the field degrees of freedom
~\cite{Ozo04}. This representation is obtained as a double Fourier transform
of the Wigner function~\cite{Case08}. For each atomic block, we define:
\begin{equation}\label{chf}
w_{ij}(k,s,t)
=
\int \rmd q\,\rme^{\rmi q k}
\langle q+s/2|\varrho_{ij}(t)|q-s/2\rangle,
\ i,j=e,g .
\end{equation}
Equivalently, the characteristic function of the composite system can be written
as a $2\times 2$ matrix in atomic subspace:
\begin{equation}\label{dmatblocks}
  \rmw(\vec{r},t) =
  \left(\begin{array}{cc}
  w_{ee}(\vec{r},t) & w_{eg}(\vec{r},t)\\
  w_{ge}(\vec{r},t) & w_{gg}(\vec{r},t)
  \end{array}\right),
\end{equation}
where $\vec{r}=(k,s)$ (depending on the context, we use a column vector or row vector).

The characteristic-function description is particularly suitable for systems
with continuous field variables because it represents the dynamics in the Fourier
phase space. In this representation, the operator equations above become the
following system of \AY{linear} PDEs:
\begin{eqnarray}\nonumber
  \rmi \h{L} w_{ee}
  &=&
  \vk  (s/2+\partial_s)(w_{ge}+w_{eg})
  -\rmi\vk (k/2+\partial_k)(w_{ge}-w_{eg}),\\\nonumber
  &&\\\nonumber
  \rmi \h{L} w_{eg}
  &=&
  \Delta w_{eg}\\\nonumber
  &&+{\vk\over 2}(s-\rmi k)(w_{gg}+w_{ee})
    -\vk(\partial_s-\rmi\partial_k)(w_{ee}-w_{gg}),\\\nonumber
  &&\\
  \rmi \h{L} w_{ge}
  &=&
  -\Delta w_{ge}
  \label{wPDEs}
  \\ 
  &&+{\vk\over 2}(s+\rmi k)(w_{gg}+w_{ee})  
    -\vk(\partial_s+\rmi\partial_k)(w_{ee}-w_{gg}),\nonumber \\ \nonumber
  &&\\\nonumber
  \rmi \h{L} w_{gg}
  &=&
  \vk  (s/2-\partial_s)(w_{ge}+w_{eg})
  -\rmi\vk (k/2-\partial_k)(w_{ge}-w_{eg}),\\ \nonumber
\end{eqnarray}
where $\partial_s := \partial/\partial s$, $\partial_k := \partial/\partial k$, $\vk=g/\sqrt{2}$ and
\begin{equation}\label{L-op}
\h{L}=\partial_t+s\partial_k-k\partial_s.
\end{equation}
Part $s\partial_k-k\partial_s$ of operator $\h{L}$ generates rotations in the $(k,s)$ plane; consequently, 
radius $r=|\vec r|=\sqrt{k^2+s^2}$ is preserved along the corresponding flow.
In Fourier phase space, this rotational structure reflects the conservation of
the total number of excitations in the JCM.

It is useful to introduce the combinations
\begin{eqnarray}\nonumber
w_f := w_{ee}+w_{gg}, &\qquad&  w_x := w_{eg}+w_{ge},\\
w_y := {w_{ge}-w_{eg}\over \rmi}, &\qquad&
w_z := w_{ee}-w_{gg}. \label{wfx-yz}
\end{eqnarray}
In terms of the vector
\[
\vec{w}:=(w_f,w_x,w_y,w_z) ,
\]
the system takes the compact form
\begin{equation}\label{afceqs}
  \h{L}\vec{w}(\vec{r},t)
  =
  \mathrm{R}(\vec{r})\,\vec{w}(\vec{r},t),
\end{equation}
with
\begin{equation}
   \mathrm{R}(\vec{r}) = -\rmi \vk
   \left(
   \begin{array}{cccc}
     0 & s  &  k  & 0\\
     s  & 0 & -\rmi {\Delta/\vk} & -2\partial_s \\
     k  & \rmi {\Delta/\vk} & 0 & -2\partial_k \\
     0 & 2\partial_s  & 2\partial_k & 0
   \end{array}
   \right).
   \label{sys_w_f}
\end{equation}
The first component, $w_f(\vec r,t)$, corresponds to the characteristic function
of the reduced field state obtained after tracing over the atomic degrees of
freedom. The remaining components are associated with atomic observables.
Indeed, after tracing over the field degrees of freedom, we obtain
\begin{eqnarray}
\langle\sigma_x(t)\rangle &=& w_x(\vec{r},t)\big|_{\vec{r}=0},\\
\langle\sigma_y(t)\rangle &=& w_y(\vec{r},t)\big|_{\vec{r}=0},\\
\langle\sigma_z(t)\rangle &=& w_z(\vec{r},t)\big|_{\vec{r}=0}.
\label{SIGMA_Z}
\end{eqnarray}

For the subsequent analysis, we consider the general atom-field initial conditions
of the form
\[
  \rmw^0(\vec{r}) := \rmw(\vec{r},0),
\]
subject to the normalization and Hermiticity conditions
\begin{equation}\label{INITIAL_W0}
[\rmw^0(0)]_{11}+[\rmw^0(0)]_{22}=1,\qquad
[\rmw^0(0)]_{12}=[\rmw^0(0)]_{21}^{\ast}.
\end{equation}
Moreover, $[\rmw^0(0)]_{11}$ and $[\rmw^0(0)]_{22}$ are real.

We also consider a special case of initially factorized atom-field
states in which
\begin{equation}
\rmw^0(\vec{r}) =
\left(\begin{array}{cc}
 \varrho^0_{ee} & \varrho^0_{eg}\\
 \varrho^0_{ge} & \varrho^0_{gg}
 \end{array}\right)
 w^0_f(\vec{r}).
 \label{IN_CON_W}
\end{equation}
Here, $\varrho^0_{ij}:=\varrho_{ij}(0)=w^0_{ij}(0)$, 
$w_f^0(\vec r):=w_f(\vec r,0)$, while the reduced atomic density-matrix elements are given by:
\begin{equation}
\varrho_{ij}(t):=w_{ij}(\vec{r},t)\big|_{\vec r=0}.
\end{equation}
The reduced atomic density-matrix elements must satisfy
\begin{eqnarray}\nonumber
\varrho_{ee}(t)+\varrho_{gg}(t)=1,\
\varrho_{ee}(t),\varrho_{gg}(t)\in\mathbb{R},\
\varrho_{eg}(t)=\varrho_{ge}^{\ast}(t).\\\label{atom_rho_cond}
\end{eqnarray}
In addition, the reduced field characteristic function $w_f(\vec r,t)$ must
satisfy the quantum-mechanical admissibility conditions listed in
Appendix~\ref{app1}.

Finally, because operator $\h{L}$ has a natural rotational structure, it is
convenient to introduce polar coordinates in the Fourier phase space:
\[
s=r\cos\varphi,\qquad k=r\sin\varphi .
\]
In these variables,
\begin{equation}\label{L-op-polar}
\h{L}=\partial_t+\partial_\varphi .
\end{equation}
We then perform the transformation
\begin{equation}\label{rotpt}
  \vec{w}=\mathrm{T}(r,\varphi)\vec{u},
\end{equation}
where
\begin{equation}
   \mathrm{T}(r,\varphi) =
   \left(
   \begin{array}{cccc}
     1 & 0  &  0 & 0\\
     0 &  \rmi\cos(\varphi)/r  & -\rmi \sin(\varphi)/r & 0 \\
     0 &  \rmi\sin(\varphi)/r &   \rmi \cos(\varphi)/r & 0 \\
     0 & 0  & 0  & 1
   \end{array}
   \right),
\end{equation}
and $\vec u=(u^1,u^2,u^3,u^4) $. The \AY{linear} system for $\vec u$ becomes
\begin{equation}\label{eqs}
   \h{L}\vec{u}=\mathrm{S}(\vec{r})\vec{u},
\end{equation}
with
\begin{equation}
   \mathrm{S}(\vec{r}) =
   \left(
   \begin{array}{cccc}
     0 & \vk   &  0 & 0\\
     -\vk r^2  & 0 & -\delta & 2\vk r\partial_r \\
     0  & \delta & 0 & 2\vk\partial_{\varphi} \\
     0 & {2\vk \over r}\partial_r & {2\vk \over r^2}\partial_{\varphi} & 0
   \end{array}
   \right).
\end{equation}
This transformed system is the starting point for the derivation of the symmetries, 
invariant solutions, and conservation laws in the following sections.

\section{\label{LST} Symmetries}
In this section, we introduce the symmetry framework used to analyze the
differential system obtained for the JCM. We follow the formulation in terms of
generating functions of symmetries~\cite{Vinogradov:1984aa,Vinogradov1989_1,Vinogradov1999}.
Let
\begin{equation}\label{PDEsystem}
\vec{F}=\left(F^1,\ldots,F^m\right) =0
\end{equation}
be a determined regular system of PDEs of order $d$, where each equation has the form
\[
F^\alpha =
F^\alpha\left(\vec{x},\vec{u},\vec{u}^{(1)},\ldots,\vec{u}^{(d)}\right),
\qquad \alpha=1,\ldots,m
\]
where $\vec{x}=(x^1,\ldots,x^n)\in\mathbb{R}^n$ denotes the independent
variables, $\vec{u}=(u^1,\ldots,u^m) $ denotes the dependent variables, and
$\vec{u}^{(s)}$ denotes the collection of all derivatives of $\vec u$ of order
$s$, with $1\leq s\leq d$.

The universal linearization operator~\cite{Vinogradov:1984aa}, also known as the Fr\'echet derivative of
the system~\cite{Sokolov2009}, is the matrix differential operator obtained by
linearizing $\vec F$ with respect to the dependent variables and their
derivatives, denoted by
\begin{equation}\label{UnivLinOp}
l_{\vec F}=\left(\left[l_{\vec F}\right]_{\alpha\beta}\right), \quad \alpha,\beta=1,\dots,m.
\end{equation}
Its components are given by
\begin{equation}
\label{l_F}
 \left[l_{\vec{F}}\right]_{\alpha \beta}
 =
 \sum_{0\leq |\varsigma|\leq d}
 {\partial F^\alpha \over \partial u_{\varsigma}^\beta}
 D^{(\varsigma)} .
\end{equation}
In this expression, $\varsigma=(\varsigma^1,\ldots,\varsigma^n)$ is a
multi-index, understood as an ordered set of 
non-negative integers, and 
\[
|\varsigma|=\varsigma^1+\cdots+\varsigma^n,\quad D^{(\varsigma)}
=
D_1^{\varsigma^1}\circ\cdots\circ D_n^{\varsigma^n}.
\]
The notation
\[
u^\beta_\varsigma =\partial^{\varsigma^1}_{x^1}...\partial^{\varsigma^n}_{x^n} u^\beta
\]
denotes the derivative of $u^\beta$ corresponding to multi-index
$\varsigma$. Operator $D_j$ is the total derivative operator with respect to
$x^j$, expressed in expanded form as
\begin{equation}\label{TOTAL_D}
D_j
=
\partial_{x^j}
+
u^\alpha_j\partial_{u^\alpha}
+
u^\alpha_{ij}\partial_{u^\alpha_i}
+
u^\alpha_{ijk}\partial_{u^\alpha_{ik}}
+\cdots ,
\end{equation}
with summation over repeated indices.

A function
\[
\vec{\phi}\left(\vec{x},\vec{u},\vec{u}^{(1)},\ldots\right)
=
(\phi^1,\ldots,\phi^m) 
\]
is called the generating function, or characteristic~\cite{Olver1986}, of a
symmetry admitted by the system $\vec{F}=0$, if it is a solution of the following system
\begin{equation}\label{l_F_phi}
\left. l_{\vec{F}}\vec{\phi} \right|_{\mathcal{F}=0}=0.
\end{equation}
Here, $\mathcal{F}=0$ denotes the infinite prolongation of the system
$\vec{F}=0$, that is, the system together with all its differential
consequences
\[
D^{(\varsigma)}F^\alpha=0.
\]
The order of the generating function is determined by the highest order of the 
derivatives of $\vec u$ appearing in $\vec{\phi}$.

In the present problem, the independent variables are
\[
x^1=t,\qquad x^2=r,\qquad x^3=\varphi,
\]
and the dependent variables are
\[
\vec u=(u^1,u^2,u^3,u^4)  .
\]
The infinitesimal operator of symmetry $X$ has the form
\begin{equation}
    X=\xi^i\partial_{x^i}+\eta^{\alpha}\partial_{u^{\alpha}},
    \qquad i=1,2,3,\quad \alpha=1,\ldots,4.
    \label{X}
\end{equation}
The corresponding generating function is related to the coefficients of $X$ by
\begin{equation}\label{phi_to_X}
\phi^\alpha = \eta^\alpha - \xi^i u^\alpha_i.
\end{equation}
If the coefficients $\xi^i$ and $\eta^\alpha$ depend only on variables
$x^i$ and $u^\alpha$, there is a point symmetry. If they depend
also on the derivatives of $\vec u$, the corresponding symmetry is a generalized symmetry.


For the JCM system written in form (\ref{eqs}), we now search for  first-order
generating functions,
\begin{equation}\label{phi}
\vec{\phi}=\vec{\phi}\left(\vec{x},\vec{u},\vec{u}^{(1)}\right),
\end{equation}
satisfying the condition (\ref{l_F_phi}). For convenience,
we introduce the radial variable
\begin{equation}\label{R}
R=\frac{r^2}{4},
\end{equation}
\AY{related with coordinates $s = 2\sqrt{R} \cos\varphi$, $k = 2\sqrt{R} \sin\varphi$.}

This simplifies the radial part of the differential operators, and any further reference to the 
capital letter \(R\) refers to this variable. For this
variable, the universal linearization operator associated with system
(\ref{eqs}) is as follows:
\begin{equation}\label{l_F_JM}
l_{\vec{F}} =
\begin{pmatrix}
D_{t}+D_{\varphi} & -\vk & 0 & 0\\
4\vk R & D_t+D_{\varphi} & \delta & -4\vk R D_R\\
0 & - \delta & D_{t}+D_{\varphi} & -2\vk D_{\varphi}\\
0 & - \vk D_R & -\frac{\vk}{2R}D_{\varphi} & D_t+D_{\varphi}
\end{pmatrix}.
\end{equation}

Solving Eq.~(\ref{l_F_phi}) for the first-order generating functions yields the
following five independent generating functions:
\begin{eqnarray}
\label{gen_function_phi1}
\vec{\phi}_1
&=&
\left(\partial_t u^1,\partial_t u^2,\partial_t u^3,\partial_t u^4\right) ,
\\
\label{gen_function_phi2}
\vec{\phi}_2
&=&
\left(\partial_\varphi u^1,\partial_\varphi u^2,
\partial_\varphi u^3,\partial_\varphi u^4\right) ,
\\
\label{gen_function_phi3}
\vec{\phi}_3
&=&
\left(u^1,u^2,u^3,u^4\right) ,
\\
\label{gen_function_phi4}
\vec{\phi}_4
&=&
\left(\tilde{u}^1,\tilde{u}^2,\tilde{u}^3,\tilde{u}^4\right) ,
\\
\label{gen_function_phi5}
\vec{\phi}_5
&=&
\left(
\frac{\delta}{\vk}u^4+\frac{\partial_{\varphi}u^2}{2R}
-\partial_R u^3,
\,
-2\partial_\varphi u^1,
\right.
\nonumber\\
&&
\left.
4R\left(-u^4+\partial_R u^1\right),
\,
\frac{\delta}{\vk}u^1-u^3
\right)  .
\end{eqnarray}
Here, $\vec{\tilde u}(t,r,\varphi)=(\tilde u^1,\tilde u^2,\tilde u^3,\tilde u^4) $
denotes an arbitrary solution of the same linear system (\ref{eqs}). 

The first four functions (\ref{gen_function_phi1})--(\ref{gen_function_phi4}) represent point symmetries (see the next subsection). The generating function $\vec{\phi}_5$ does not correspond to the point symmetry.
This is a first-order generalized symmetry with operator 
$X^{(1)}=\phi_5^\alpha\partial_{u^\alpha}$, whose coefficients $\eta^\alpha$ depend on the first derivatives of $\vec{u}$.

Let us now consider two diagonal matrix differential operators:
\[
\mathrm{D}_t:=\operatorname{diag}(D_t,D_t,D_t,D_t),
\quad
\mathrm{D}_\varphi:=\operatorname{diag}(D_\varphi,D_\varphi,D_\varphi,D_\varphi).
\]
One can verify that these two operators commute with $l_{\vec F}$:
\[
\mathrm{D}_t\circ l_{\vec F}=l_{\vec F}\circ \mathrm{D}_t,
\quad
\mathrm{D}_\varphi\circ l_{\vec F}=l_{\vec F}\circ \mathrm{D}_\varphi .
\]
Consequently, if \(\vec\phi\) is a solution to
(\ref{l_F_phi}), then \(\mathrm{D}_t \vec\phi \) and
\(\mathrm{D}_\varphi \vec\phi \) are also solutions. 
These operators are known as
{\it recursion operators} since they allow us to obtain new generating 
functions of symmetry from the known generating functions. In particular, repeated applications of \(\mathrm{D}_t\) and
\(\mathrm{D}_\varphi\) to the generating functions
\(\vec\phi_3\) ($\vec\phi_1=\mathrm{D}_t\vec\phi_3$, $\vec\phi_2=\mathrm{D}_\varphi\vec\phi_3$) and \(\vec\phi_5\) produce generalized
symmetries of arbitrarily high order. More precisely, if \(\vec\phi\) is a
generating function of order \(q\), then \(\mathrm{D}_t^k\vec\phi\) and \(\mathrm{D}_\varphi^k\vec\phi\) are the generating functions of order \(q+k\). This shows that the JCM differential system admits generalized symmetries of an arbitrary order.

We also note that the direct calculation of second-order generating functions does not yield additional
independent symmetries beyond those obtained using these two recursion
operators. \AY{These calculations are quite cumbersome. We used the symbolic computation system Maple 2020.1 (licensed to Universidad de Guadalajara) for verification (see Ref. \onlinecite{Pinuelas:2026}).}

\subsection{\label{sec:point_symmetries}Point symmetries and invariant solutions}

In Lie symmetry theory, point symmetries admitted by a system of differential
equations can be used to construct classes of invariant solutions. These 
solutions remain unchanged under the action of the selected symmetry group.
After substituting the corresponding invariant ansatz into the original system,
one obtains a reduced system involving fewer independent variables or
simpler dependence on the original variables. This reduced system, often called
a factor-system, is typically more analytically accessible and provides a
systematic route for deriving exact solutions.

Using the relation between infinitesimal generators and generating functions,
Eq.~(\ref{phi_to_X}), applied to the first four generating functions in
Eqs.~(\ref{gen_function_phi1})--(\ref{gen_function_phi4}), we obtain the point
symmetry generators \AY{(given that the basis elements allow for a sign change, we consider $-X_i \mapsto X_i$)} 
\begin{equation}
X_1=\partial_t,\quad
X_2=\partial_\varphi,\quad
X_3=u^\alpha\partial_{u^\alpha},\quad
X_4=\tilde{u}^\alpha\partial_{u^\alpha}.
\end{equation}
Generators $X_1$ and $X_2$ correspond to translations in $t$
and $\varphi$, respectively, while $X_3$ generates a scaling transformation of
the dependent variables $u^\alpha$. \AY{The set of operators of the form $X_4$ constitutes an infinite-dimensional abelian Lie algebra $L^\infty$ that is an ideal in the algebra of point symmetries admitted by the system of linear equations\cite{Ovsiannikov82}. This fact represents the superposition of the solutions of the linear system of PDEs. Thus, the Lie algebra of point symmetries $L_3 \oplus L^\infty$ consists of the Abelian three-dimensional subalgebra $L_3 = \langle X_1, X_2, X_3 \rangle$ and the ideal $L^\infty$.}

The set of non-equivalent one-dimensional subalgebras of $L_3$ is given by\cite{Patera:1977aa}:
\begin{equation}\label{algebra}
\Theta_1=\left\langle X_2 \right\rangle,\
\Theta_2=\left\langle X_1+\alpha_0 X_2 \right\rangle,\
\Theta_3=\left\langle X_3+\alpha_0 X_1+\beta_0 X_2 \right\rangle,
\end{equation}
where $\alpha_0$ and $\beta_0$ are arbitrary constants.


\subsubsection{\label{subsec:Inv1}Invariant solution for $\Theta_1$}

To determine the invariant solutions associated with each subalgebra, we first
find the invariants $J$ of the corresponding basis operators. For the subalgebra
$\Theta_1=\langle X_2\rangle$, the invariance condition
$X_2J=0$ gives $J_\alpha=u^\alpha,$ $J_5=t$, $J_6=r$. 
Therefore, the invariant solution has the form
\begin{equation}
u^\alpha(\vec r,t)=f^\alpha(r,t),
\end{equation}
that is, it has no explicit dependence on the angular variable $\varphi$.
Substitution of this ansatz into the dynamical system (\ref{eqs}) gives the
factor-system
\begin{eqnarray}
\partial_t f^1 &=& \vk f^2, \nonumber\\
\partial_t f^2 &=& -\delta f^3-\vk r\left(r f^1-2\partial_r f^4\right),
\nonumber\\
\partial_t f^3 &=& \delta f^2, \nonumber\\
\partial_t f^4 &=& {2\vk\over r}\partial_r f^2.
\label{eff1}
\end{eqnarray}
The equation for $f^2$ can be decoupled by taking the second derivative with
respect to $t$. Using the remaining equations of the factor-system, one obtains
\begin{equation}
    \partial_t^2 f^2+\delta^2 f^2
    =
    4\vk^2 R\left(\partial_R^2 f^2-f^2\right).
\end{equation}
We now use separation of variables, $f^2=T(t)Y(R)$.
Substitution into the previous equation leads to
\begin{eqnarray}
    T^{\prime\prime}+(\delta^2+4\vk^2\ell)T&=&0,
    \label{inv1vsep}\\
    Y^{\prime\prime}-\left(1-\frac{\ell}{R}\right)Y&=&0,
    \label{Y}
\end{eqnarray}
where $-4\vk^2\ell$, with $\ell\neq 0$, has been chosen as the separation
constant.

By introducing
\[
Y(R)=R\rme^{-R}y(\tilde{x}),\qquad \tilde{x}=2R,
\]
Eq.~(\ref{Y}) becomes Kummer's equation,
\begin{equation}
\tilde{x}y^{\prime\prime}+(2-\tilde{x})y^\prime
+\left(\frac{\ell}{2}-1\right)y=0.
\end{equation}
Its general solution is~\cite{BATEMAN:1}
\begin{equation}
y(\tilde{x})
=
c_1\,{}_1F_1(1-\ell/2;2;\tilde{x})
+
c_2\,U(1-\ell/2;2;\tilde{x}),
\end{equation}
where ${}_1F_1$ is the confluent hypergeometric function of the first kind and
$U$ is the confluent hypergeometric function of the second kind. The asymptotic
behavior of these functions as $R\to\infty$ is given by
(see Eqs.~13.2.23 and 13.2.6 of Ref.~\onlinecite{OLVER_F})
\begin{eqnarray}
{}_1F_1(a;b;z)
&\sim&
\frac{\rme^z z^{a-b}\Gamma(b)}{\Gamma(a)},
\qquad a\neq 0,-1,\ldots,\\
U(a,b,z)
&\sim&
z^{-a}.
\end{eqnarray}
Thus, one must either set $c_1=0$ or impose
\[
1-\frac{\ell}{2}=-n,\qquad n=0,1,\ldots .
\]
In the first case, the function $U(1-\ell/2;2;2R)$ is one-valued
\cite{BATEMAN:1} only if either (i) $1-\ell/2=-n$, in which case it is a
polynomial in $R$ and is proportional to ${}_1F_1$, or (ii)
$1-\ell/2=1$, corresponding to $\ell=0$. Therefore, the only one-valued
solution that is compatible with the finiteness and integrability of the chord
function is proportional to:
\[
R\rme^{-R}\,{}_1F_1(-n;2;2R),
\qquad
\ell=2(n+1).
\]

Using the relation between confluent hypergeometric functions and associated
Laguerre polynomials,
\begin{equation}\label{hyptoLag}
    L^{(m)}_n(x)
    =
    \frac{(m+n)!}{m!n!}\,{}_1F_1(-n;m+1;x),
\end{equation}
the radial solution can be written, up to an overall constant, as
\begin{equation}
 Y_n(r):=r^2\rme^{-r^2/4}L^{(1)}_n(r^2/2).
 \label{Y_n}
\end{equation}
Furthermore, using the identity
\begin{equation}
  L^{(1)}_n(x)
  =
  {n+1\over x}\left[L_n(x)-L_{n+1}(x)\right],
\end{equation}
we can equivalently write
\begin{equation}
    Y_n(r)
    =
    2(n+1)\rme^{-r^2/4}
    \left[
    L_n(r^2/2)-L_{n+1}(r^2/2)
    \right].
    \label{Y_n_alt}
\end{equation}

From Eq.~(\ref{inv1vsep}), together with the quantization condition
$\ell=2(n+1)$, the time-dependent part becomes
\begin{equation}
    T^+_n(t)
    :=
    \tilde c_1\rme^{\rmi\lambda_n t}
    +
    \tilde c_2\rme^{-\rmi\lambda_n t},
\end{equation}
where $\lambda_n$ is the generalized Rabi frequency defined in
Eq.~(\ref{lambn}). The time-dependent functions $T^+_n(t)$ and $T^-_n(t)$,
defined below, reproduce the same oscillatory structure as the dressed-state
solution in Eqs.~(\ref{soldress})--(\ref{psit}), encoded in amplitudes
$a_n(t)$ and $b_n(t)$ in Eq.~(\ref{atbt}), for appropriate initial conditions
(see Appendix~\ref{appJCM}).

Thus, the invariant solutions $\vec u_n$ obtained from $\Theta_1$ and
associated with frequency $\lambda_n$ are
\begin{eqnarray}
  u^1_n(\vec r,t)
  &=&
  -{\rmi g\over \lambda_n\sqrt{2}}\,Y_n(r)T^-_n(t)+U^1(r),
  \label{sol11}\\
  u^2_n(\vec r,t)
  &=&
  Y_n(r)T^+_n(t),\\
  u^3_n(\vec r,t)
  &=&
  -{\rmi\delta\over\lambda_n}\,Y_n(r)T^-_n(t)+U^3(r),\\
  u^4_n(\vec r,t)
  &=&
  -{\rmi g\sqrt{2}\over\lambda_n}\,
  {Y_n^\prime(r)\over r}T^-_n(t)+U^4(r),
  \label{sol14}
\end{eqnarray}
where
\begin{eqnarray}
T^-_n(t)
&:=&
\tilde c_1\rme^{\rmi\lambda_n t}
-
\tilde c_2\rme^{-\rmi\lambda_n t},
\label{relarbfr}\\
U^1(r)
&:=&
{2\over r}U^{4\,\prime}(r)
-
{\delta\sqrt{2}\over gr^2}U^3(r),
\label{UUU}
\end{eqnarray}
and $U^3(r)$ and $U^4(r)$ are arbitrary functions \AY{such that $U^{1,3,4}(0)$ are defined}.

Using the inverse transformation between the variables $u^\alpha$ and the
characteristic-function components, we obtain: 
\begin{eqnarray}
w_{ee}&=\!\!&\!\!
{u^1_n(\vec r,t)+u^4_n(\vec r,t)\over 2},
\!\!\!\quad\!\!\!
w_{gg}
\!=\!
{u^1_n(\vec r,t)-u^4_n(\vec r,t)\over 2},
\\
w_{eg}
&=&
(k+\rmi s){u^2_n(\vec r,t)\over 2r^2}
+
(s-\rmi k){u^3_n(\vec r,t)\over 2r^2},
\label{EQ_w_eg}\\
w_{ge}
&=&
(-k+\rmi s){u^2_n(\vec r,t)\over 2r^2} - (s+\rmi k){u^3_n(\vec r,t)\over 2r^2}.
\label{EQ_w_ge}
\end{eqnarray}
\AY{For the convenience of the reader, we present the inverse transformations for (\ref{wfx-yz}) and (\ref{rotpt}):
\bea
&u^1 = w_f, u^2 = -\rmi\left(s w_x + k w_y\right),  u^3 = \rmi\left(k w_x - s w_y\right),  u^4 = w_z; \nonumber\\
&w_{ee} = \frac{w_f + w_z}{2}, w_{gg} = \frac{w_f - w_z}{2}, w_{eg} = \frac{w_x + \rmi w_y}{2}, w_{ge} = \frac{w_x - \rmi w_y}{2}. \nonumber
\eea
}

One can then verify that the initial conditions (\ref{INITIAL_W0}) and the
conditions a), b), and c) of Appendix~\ref{app1} for the characteristic
function $w_f(\vec r,t)$ are satisfied if $a_1$ and $b_1$ are real constants,
$U^1(r),U^3(r),U^4(r)\in\mathbb{R}$, $U^1(r)$ is bounded, $U^1(0)=1$, and
\begin{eqnarray}
T^+_n(t)
&=&
a_1\cos\lambda_n t-b_1\sin\lambda_n t,\\
T^-_n(t)
&=&
\rmi\left(a_1\sin\lambda_n t+b_1\cos\lambda_n t\right).
\end{eqnarray}

The condition of the initially decoupled atom--field states, Eq.~(\ref{IN_CON_W}),
together with the boundary conditions and the conditions a), b), and c) of
Appendix~\ref{app1}, are fulfilled by choosing
\begin{eqnarray}
  u^1_n(\vec r,t)
  &=&
  {cg\over \lambda_n\sqrt{2}}\,Y_n(r)\cos\lambda_n t
  +U^1_n(r),
  \label{sol112}\\
  u^2_n(\vec r,t)
  &=&
  -cY_n(r)\sin\lambda_n t,\\
  u^3_n(\vec r,t)
  &=&
  {c\delta\over\lambda_n}\,Y_n(r)(\cos\lambda_n t-1),\\
  u^4_n(\vec r,t)
  &=&
  {cg\sqrt{2}\over\lambda_n}
  {Y_n^\prime(r)\over r}(\cos\lambda_n t-1),
  \label{sol141}
\end{eqnarray}
where
\begin{eqnarray}
 T^+_n(t)&=&-c\sin\lambda_n t,
 \qquad
 T^-_n(t)=\rmi c\cos\lambda_n t,\\
 U^3_n(r)&=&
 -{c\delta\over\lambda_n}Y_n(r),
 \qquad
 U^4_n(r)=
 -{cg\sqrt{2}\over\lambda_n}{Y'_n(r)\over r},
\end{eqnarray}
with
\[
c:={\vk\over (n+1)\lambda_n}.
\]
In this case, the reduced field characteristic function
$w_f(\vec r,t)=u^1_n(\vec r,t)$ takes the explicit form
\begin{equation}
w^{(n)}_f(\vec r,t)
=
{cg(\cos\lambda_n t-1)\over\lambda_n\sqrt{2}}Y_n(r)
+
{c\lambda_n\sqrt{2}\over g}{Y_n(r)\over r^2}.
\label{w_f_1}
\end{equation}
Moreover, $\rho_{eg}(t)=\rho_{ge}(t)\equiv 0$, while
$w_z(\vec r,t)=u^4_n(\vec r,t)$. To simplify the expression for
$U^1_n(r)$, we use the relation
\begin{equation}
{Y^\prime_n(r)\over r}
=
{Y_n(r)\over 2}
+
2(n+1)\rme^{-r^2/4}L_{n+1}(r^2/2).
\label{Y'_r2}
\end{equation}
\AY{
Note that all functions (\ref{sol112})--(\ref{sol141}) are defined at the point $r = 0$ in view of Eqs. (\ref{Y_n}) and (\ref{Y'_r2}).}

\subsubsection{\label{subsec:Inv2}Invariant solution for $\Theta_2$}
We now consider the invariant solutions associated with subalgebra
$\Theta_2$ in Eq.~(\ref{algebra}). Because this subalgebra depends on the  
parameter $\alpha_0$, different choices of $\alpha_0$ lead to different
classes of invariant solutions. The invariants are obtained from
\begin{equation}
    \left(\partial_t+\alpha_0\partial_{\varphi}\right)J=0.
\end{equation}
Thus,
\[
J_\alpha=u^\alpha,\qquad
J_5=\varphi-\alpha_0 t=:\theta,\qquad
J_6=r,
\]
and the corresponding invariant solution has the form
\[
u^\alpha(\vec r,t)=f^\alpha(\theta,r).
\]

For a particular value $\alpha_0=1$, one has $\hat L f^\alpha=0$, and 
system (\ref{eqs}) leads to the stationary part of the solution
(\ref{sol11})--(\ref{sol14}), obtained by setting $T_n^\pm\equiv 0$.
Therefore, we focus on the nontrivial case
\[
\alpha_0=1+\frac{\vk}{a},\qquad a\in\mathbb{R}\setminus\{0\}.
\]
In this case, the invariant ansatz yields the factor-system
\begin{eqnarray}
\partial_\theta f^1
&=&
-a f^2,
\label{fa122}\\
\partial_\theta f^2
&=&
{a\delta\over\vk}f^3+4aR\left(f^1-\partial_R f^4\right),
\label{fa222}\\
\partial_\theta f^3
&=&
-{a\delta\over\vk}f^2-2a\partial_\theta f^4,
\label{fa322}\\
\partial_\theta f^4
&=&
-a\partial_R f^2-{a\over 2R}\partial_\theta f^3,
\label{fa422}
\end{eqnarray}
where, as before, $R=r^2/4$.

After eliminating $f^1$, $f^3$, and $f^4$ from
Eqs.~(\ref{fa122})--(\ref{fa422}), one obtains a closed equation for $f^2$ as follows:
\begin{equation}\label{ECV22}
\partial_\theta^2 f^2
=
B(R)\partial_R^2 f^2
+
C(R)\partial_R f^2
+
D(R)f^2,
\end{equation}
where
\begin{eqnarray}
B(R)
&=&
{4(aR)^2\over R-a^2},
\nonumber\\
C(R)
&=&
-{4a^4R\over (R-a^2)^2},
\nonumber\\
D(R)
&=&
-{4a^2R\over (R-a^2)^2}
\left[
(R-a^2)^2
+
{\delta^2\over 4\vk^2}(R-a^2)
-
{a\delta\over 2\vk}
\right].\nonumber\\
\end{eqnarray}

We now use the separation of variables $f^2=T(\theta)Y(R)$.
Substitution into Eq.~(\ref{ECV22}) gives
\begin{eqnarray}
{T^{\prime\prime}\over T}
&=&
B(R){Y^{\prime\prime}\over Y}
+
C(R){Y^\prime\over Y}
+
D(R) = \ell,
\label{ECV22b}
\end{eqnarray}
where $\ell$ is the separation constant. Hence, the radial function satisfies
\begin{equation}\label{Ydeq}
  B(R)Y^{\prime\prime}+C(R)Y^\prime+\left[D(R)-\ell\right]Y=0.
\end{equation}
With the change of variables
\[
Y(R)=\rme^{-R}R^{\sqrt{-\ell}/2}h(z),
\qquad
z={R\over a^2},
\]
Eq.~(\ref{Ydeq}) is transformed into the confluent Heun equation
\begin{equation}\label{C_HEUN}
h^{\prime\prime}
+
\left(
\alpha_h
+
{\beta_h+1\over z}
+
{\gamma_h+1\over z-1}
\right)h^\prime
+
\left(
{\mu_h\over 2z}
+
{\nu_h\over 2(z-1)}
\right)h
=0,
\end{equation}
where
\bea
\mu_h
&=&
\alpha_h\beta_h-\beta_h\gamma_h+\alpha_h-\beta_h-\gamma_h-2\eta_h, \nonumber
\\
\nu_h
&=&
\alpha_h\gamma_h+\beta_h\gamma_h+\alpha_h+\beta_h+\gamma_h
+2\delta_h+2\eta_h, \nonumber
\eea
and
\begin{eqnarray}
\alpha_h&=&-2a^2,
\qquad
\beta_h=\sqrt{-\ell},
\qquad
\gamma_h=-2,
\nonumber\\
\delta_h
&=&
{(4a^4-\ell)\vk^2-a^2\delta^2\over 4\vk^2},
\quad
\eta_h
=
1-\delta_h+{a\delta\over 2\vk}.
\label{HH}
\end{eqnarray}

Equation (\ref{C_HEUN})  has a unique particular solution, that is
regular around the regular singular point $z = 0$,  given by
\begin{equation}\label{H_SERIES}
h(z) =  \mathrm{HeunC}\left(\alpha_h, \beta_h, \gamma_h, \delta_h, \eta_h; z\right) = \sum\limits_{k=0}^\infty v_k z^k,
\end{equation}
This series is convergent in disk $|z|<1$, and the coefficients $v_k$ are
determined using a three-term recurrence relation (see Ref. \onlinecite{Fiziev_2010}). 
To obtain a convergent solution on the entire interval $z\in[0,\infty)$, one can 
consider a polynomial solution, the necessary condition of which is 
\begin{equation}
-{\delta_h\over\alpha_h}
=
N_h+1+{\beta_h+\gamma_h\over 2},
\qquad
N_h\in\mathbb{N}.
\label{NC}
\end{equation}
Using the parameter values in Eq.~(\ref{HH}), this condition gives the
quantization of the separation constant,
\begin{equation}
\ell_n
=
-2{a^2\over g^2}
\left(\lambda_n\pm ag\sqrt{2}\right)^2,
\label{ell_n}
\end{equation}
where $\lambda_n$ is defined by Eq.~(\ref{lambn}) and $N_h=n+1$.
Moreover, for $\ell=\ell_n$, the coefficients in series
(\ref{H_SERIES}) vanish from $v_{n+2}$ onward (see Appendix \ref{app:C}). Therefore,
\begin{eqnarray}
h_{n+1}(R)
&:=&
\mathrm{HeunC}\left(
-2a^2,\sqrt{-\ell_n},-2,
a^4-{\ell_n\over 4}-{a^2\delta^2\over 2g^2},
\right.
\nonumber\\
&&
\left.
1-a^4+{\ell_n\over 4}+{a^2\delta^2\over 2g^2}
+{a\delta\over g\sqrt{2}},
{R\over a^2}
\right)
\label{h_n+1}
\end{eqnarray}
is a polynomial of degree $n+1$ in $R$, with $h_{n+1}(0)=1$. Consequently, the
radial part of $f^2(\theta,r)$ is
\begin{equation}
Y_n(r)=\rme^{-r^2/4}r^{\sqrt{-\ell_n}}h_{n+1}(r^2/4).
\label{Rpart2invsol}
\end{equation}

The temporal part is obtained directly from Eq.~(\ref{ECV22b}) as
\begin{equation}\label{Tpart2invsol}
T_n^\pm(\theta)
=
\tilde c_1\rme^{\rmi\sqrt{|\ell_n|}\theta}
\pm
\tilde c_2\rme^{-\rmi\sqrt{|\ell_n|}\theta}.
\end{equation}
The corresponding invariant solution is
\begin{eqnarray}\label{uinv2-1}
 u^1_n(\vec{r}, t ) &=&\frac{\rmi a}{\sqrt{|\ell_n|}}\, Y_n(r) T^{-}_n(\theta) + U^1(r),\\
 u^2_n(\vec{r}, t ) &=&  Y_n(r) T^+_n(\theta),\\
 u^3_n(\vec{r}, t ) &=&  \frac{\rmi a}{\sqrt{|\ell_n|}} \frac{r}{r^2 - 4a^2}\left[{\delta\sqrt{2}}r Y_n(r)/{g} \right. \nonumber \\  
  &-&  \left. 4aY_n^\prime(r) \right]  T^{-}_n(\theta) + U^3(r),
  \label{uinv2-44}
  \\
 u^4_n(\vec{r}, t ) &=& \frac{2\rmi a}{\sqrt{|\ell_n|}} \frac{1}{r^2 - 4a^2}\left[-{a\delta\sqrt{2}} Y_n(r)/{g} \right. \nonumber \\  
  &+& \left. rY_n^\prime(r) \right]  T^{-}_n(\theta) + U^4(r),\label{uinv2-4}
\end{eqnarray}
where $U^1(r)$, $U^3(r)$, and $U^4(r)$ are the same arbitrary functions as those in
Eq.~(\ref{UUU}). \AY{Note that the functions $u^{3,4}_n(\vec{r}, t)$ do not have a singularity at $r = 2|a|$ (see Appendix \ref{app:C}).} 

It is useful to rewrite $T_n^\pm(\theta)$ in terms of original variables.
Because $\theta=\varphi-\alpha_0 t$ and
$s \pm \rmi k = r\rme^{\pm \rmi\varphi}$, we have
\begin{eqnarray}
T_n^\pm(\theta)
&=&
\left[
\tilde c_1
(s+\rmi k)^{\sqrt{|\ell_n|}}
\rme^{-\rmi\alpha_0\sqrt{|\ell_n|}t}
\right.
\nonumber\\
&&
\left.
\pm
\tilde c_2
(s-\rmi k)^{\sqrt{|\ell_n|}}
\rme^{\rmi\alpha_0\sqrt{|\ell_n|}t}
\right]
r^{-\sqrt{|\ell_n|}} .
\end{eqnarray}
The normalization condition for the reduced field characteristic function,
$w_f(0,t)=u^1_n(0,t)=1$, is fulfilled if $U^1(0)=1$. \AY{This condition imposes constraints on the choice of functions $U^{3,4}(r)$ defined by Eq (\ref{UUU}). Since the factor $r^{-\sqrt{|\ell_n|}}$ in $T_n^\pm(\theta)$ cancels with the factor in $Y_n(r)$, all functions (\ref{uinv2-1})--(\ref{uinv2-4}) are defined as $\vec{r}$ approaches zero.}

The Hermiticity condition
\[
w_f(-\vec r,t)=w_f^\ast(\vec r,t)
\]
leads to two possible cases. In the even case, with $m\in\mathbb{N}$, \AY{ we set $\tilde c_2=\tilde c_1^\ast$ and 
\begin{equation}
a = a_{2m} :=
{\sqrt{\lambda_n^2+8g^2m} \mp \lambda_n\over 2\sqrt{2}g}.
\end{equation}
Then, from (\ref{ell_n}) we obtain $\ell_n = -4m^2$.

In the odd case,
\begin{eqnarray}
\tilde c_2&=&-\tilde c_1^\ast,
\quad
\ell_n=-(2m+1)^2,
\nonumber\\
a&=&a_{2m+1}
:=
{\sqrt{\lambda_n^2+4g^2(2m+1)} \mp \lambda_n\over 2\sqrt{2}g}.
\end{eqnarray}
}
In both cases, $U^1(r)$ is taken to be real. Thus,
\begin{eqnarray}
w_f^{\rm even}(\vec r,t)
&:=&
\rmi {a_{2m}\over 2m}
\rme^{-r^2/4}
h_{n+1}(r^2/4)
\nonumber\\
&&\times
\left[
\tilde c_1
(s+\rmi k)^{2m}
\rme^{-\rmi 2m\alpha_0 t}
-
\tilde c_1^\ast
(s-\rmi k)^{2m}
\rme^{\rmi 2m\alpha_0 t}
\right]
\nonumber\\
&&
+U^1(r)
\label{w_f_2}
\end{eqnarray}
is real-valued, whereas
\begin{eqnarray}
w_f^{\rm odd}(\vec r,t)
&:=&
\rmi {a_{2m+1}\over 2m+1}
\rme^{-r^2/4}
h_{n+1}(r^2/4)
\nonumber\\
&&\times
\left[
\tilde c_1
(s+\rmi k)^{2m+1}
\rme^{-\rmi(2m+1)\alpha_0 t}
\right.
\nonumber\\
&&
\left.
+
\tilde c_1^\ast
(s-\rmi k)^{2m+1}
\rme^{\rmi(2m+1)\alpha_0 t}
\right]
+U^1(r)\nonumber\\
\end{eqnarray}
is, in general, complex-valued.

To satisfy conditions (\ref{INITIAL_W0}), we take $U^4(r)\in\mathbb{R}$
and restrict our attention to the even case. In the odd case, the requirement that
$[\rmw^0(\vec r)]_{11}$ and $[\rmw^0(\vec r)]_{22}$ be real forces
$\tilde c_1=0$, which reduces $w_f^{\rm odd}$ to a static characteristic
function. Furthermore, the Hermiticity condition at the origin,
\[
[\rmw^0(0)]_{12}=[\rmw^0(0)]_{21}^\ast,
\]
implies, in the limit $s,k\to0$, that
\[
[\rmw^0(0)]_{12}=[\rmw^0(0)]_{21}^\ast=0.
\]

Finally, considering the more restrictive condition of the initially decoupled
atom--field states, Eq.~(\ref{IN_CON_W}). At the limit $\vec r\to0$, one obtains
$\rho_{eg}=\rho_{ge}=0$. Hence, for a factorized initial state,
$[\rmw^0(\vec r)]_{12}$ and $[\rmw^0(\vec r)]_{21}$ must vanish identically.
Using Eqs.~(\ref{EQ_w_eg}) and (\ref{EQ_w_ge}), it is necessary that 
\[
u^2_n(\vec r,0)\equiv 0,
\qquad
u^3_n(\vec r,0)\equiv 0,
\]
which is only possible if $\tilde c_1=0$. Therefore, in the factorized case, the
solution becomes static. Consequently, the nontrivial even invariant solution
obtained from $\Theta_2$ describes generic coupled atom--field initial states
rather than initially decoupled ones.

\begin{figure}[htbp]
\begin{center}
\includegraphics[width=4cm]{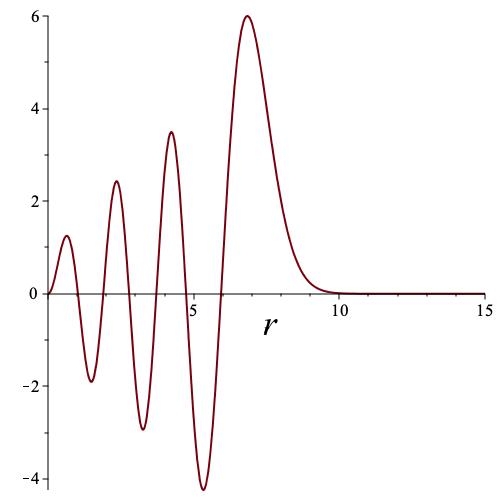}
\includegraphics[width=4cm]{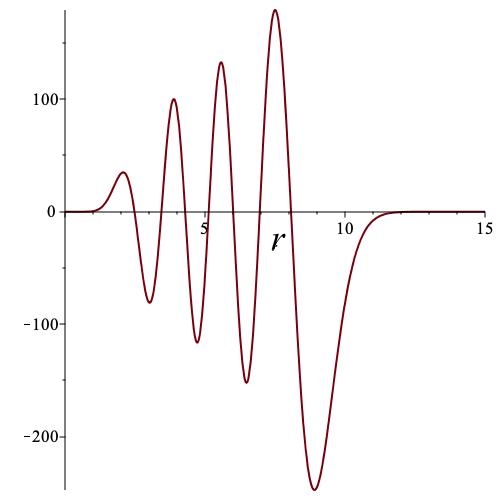}
\caption{Two radial functions $Y_6(r)$ from the first invariant solution (left) and from the second invariant solution (right).}
\label{fig:1}
\end{center}
\end{figure}

\AY{Let us compare two radial components $Y_n(r)$, expressed in terms of Laguerre polynomials (\ref{Y_n}) and Heun polynomials (\ref{Rpart2invsol}). Figure \ref{fig:1} shows these functions for the values $n = 6$, $a = \delta = g = 1$. Apart from the scale, the behavior of these functions exhibits certain similarities: a wavelike  nature, a zero at $r = 0$, and decay at infinity. However, there is a significant difference in the number of zeros: the radial function, expressed in terms of Heun polynomials, has one additional root. It should also be noted that the first radial function depends only on the number $n$, whereas the second depends also on the eigenvalues $\lambda_n$.}

\begin{figure}[htbp]
\begin{center}
\includegraphics[width=4cm]{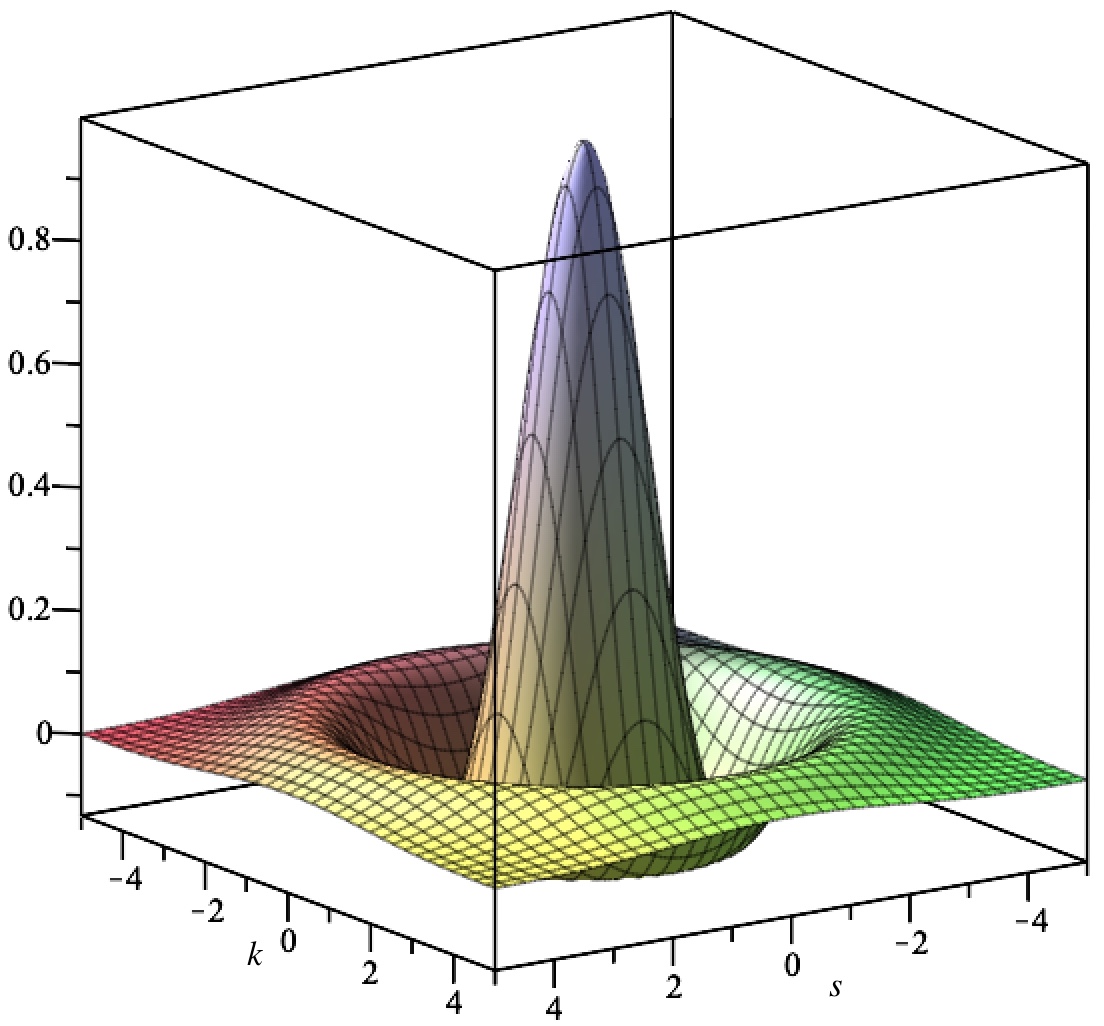}
\includegraphics[width=4cm]{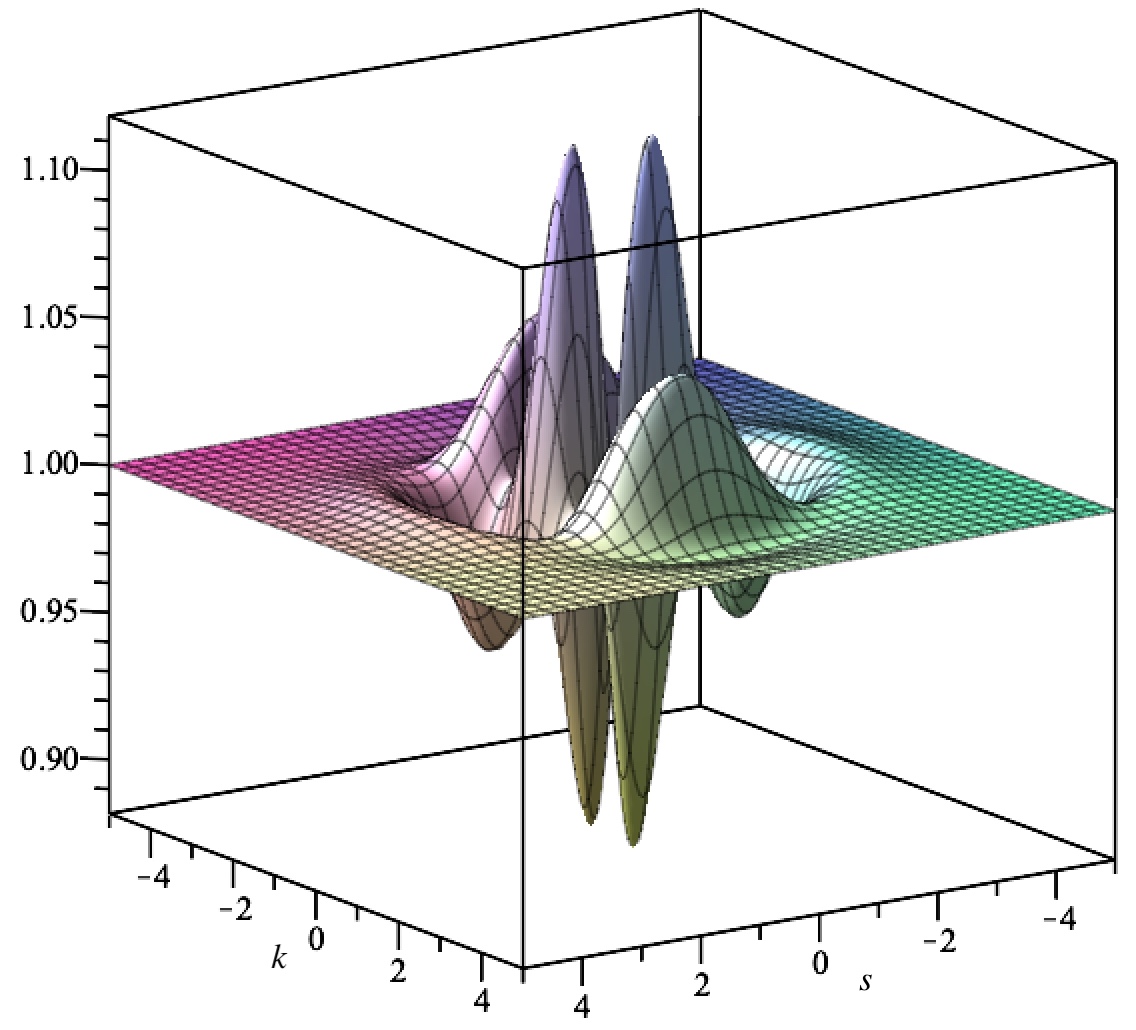}
\caption{Two functions $w_f(\vec{r},t)$ from the first invariant solution (left) and from the second invariant solution (right). Values of parameters are: $\delta = g = 1$, $n = 2$, $m = 1$, $\tilde{c}_1 = \rmi$.}
\label{fig:3}
\end{center}
\end{figure}

\AY{We also compare two characteristic functions $w_f$ for the first (\ref{w_f_1}) and second (\ref{w_f_2}) invariant solutions (in the second case, we assume $U^1(r) \equiv 1$). For a fixed $n$, just as with the radial parts, the number of roots for the second invariant solution is greater by one. Furthermore, the first function oscillates over time (while the roots remain fixed), whereas the time evolution of the second function results in rotation within the $(s, k)$ plane without oscillation. Figure \ref{fig:3} shows the two characteristic functions for $n=2$ at the initial time.
}

\AY{
Let us consider an expansion of the Heun polynomial $h_{n+1}(R)$ {(\ref{h_n+1})} in the associated Laguerre basis:
\begin{equation}
h_{n+1}(R)
=
\sum_{j=0}^{n+1}
d_{j}
L_{j}^{(2m)}(2R).
\label{eq:heun_laguerre_expansion}
\end{equation}
Notice that expanding (\ref{eq:heun_laguerre_expansion}) in powers of $R$ we obtain the 
system of linear non-homogeneous algebraic equations for unknown $d_j$ with a triangular matrix.  
Taking into account, that the coefficient of $R^{j}$ in $L_{j}^{(2m)}(2R)$ 
is equal to $(-2)^{j}/j!$, the determinant of the system is equal to 
$\prod\limits_{j = 0}^{n + 1} (-2)^{j}/j! \ne 0$. Therefore, exist unique values of $d_j$ for 
fixed numbers $n$ and $m$. 

According to Eqs. (\ref{eq:fock_element_1}), (\ref{eq:fock_element_2}),  one can identifies a finite structured superposition of Fock-space coherences connecting photon-number states separated by
$2m$ excitations. Thus, by defining  
\begin{equation}
G_{j}
:=
d_{j} \frac{2^{m}a_{2m}}{2m}
\sqrt{\frac{(j+2m)!}{j!}},
\label{eq:G_j}
\end{equation}
the Heun contribution part of the field characteristic function 
corresponds to:
\begin{equation}
\begin{aligned}
\Delta w_{f}(\vec{r},t)
:={}&
\rmi \sum_{j=0}^{n+1}
G_{j}
\Big[
\tilde{c}_{1}
\rme^{-\rmi2m\alpha_{0}t} \, \chi_{j ,j + 2m }(\vec{r}\,)
\\
&\qquad\qquad
-
\tilde{c}_{1}^{*}
\rme^{\rmi2m\alpha_{0}t} \, \chi_{j + 2m, j }(\vec{r}\,)
\Big].
\end{aligned}
\label{eq:rho_heun}
\end{equation}

Function $U^{1}(r)$ in (\ref{w_f_2}) must provide the diagonal populations required to make the Fock coherences consistent with a positive semidefinite density operator. In particular, one could take
\begin{equation}
U^{1}(r)
=
\sum_{j=0}^{n + 1}
p_{j}
\rme^{-r^{2}/4}
L_{j}\left(
{r^{2}}/{2}
\right),\quad p_{j}\geq0,
\
\sum_{j=0}^{n + 1}p_{j}=1.
\label{eq:U1_choice}
\end{equation}
Therefore, the solution for the non static part takes the form: 
\bea
w_f^{\rm even}(\vec r,t) = \Delta w_{f}(\vec{r},t) +  \sum_{j=0}^{n + 1} p_{j}\chi_{j,j}(\vec{r}).
\label{eq:full_wf_even}
\eea

The physical meaning of (\ref{eq:full_wf_even}) becomes clearer, if it is written in the density operator form: 
\begin{equation}
\varrho_{f}(t) = \varrho_{\mathrm{diag}}
+\Delta\varrho_{f}(t),\quad \varrho_{\mathrm{diag}}
= \sum_{j=0}^{n + 1} p_{j} |j\rangle\langle j|,
\label{eq:rho_total_field_diag}
\end{equation}
and 
\begin{equation}
\begin{aligned}
\Delta\varrho_{f}(t)
={}&
\rmi
\sum_{j=0}^{n+1}
G_{j}
\Big[
\tilde{c}_{1}
\rme^{-\rmi2m\alpha_{0}t}
|j+2m\rangle\langle j|
\\
&\qquad\qquad
-
\tilde{c}_{1}^{*}
\rme^{\rmi2m\alpha_{0}t}
|j\rangle\langle j+2m|
\Big].
\end{aligned}
\label{eq:rho_heun}
\end{equation}
}

\subsubsection{Invariant solution for $\Theta_3$}

Finally, we consider subalgebra $\Theta_3$. For a nontrivial invariant
solution associated with this subalgebra, it is necessary that
\[
\alpha_0^2+\beta_0^2\neq 0.
\]
We analyzed the possible cases separately and showed that none of them satisfies
the boundary condition required for the characteristic function.

First, let $\beta_0=0$. In this case, the invariant solution has the form
\[
u^\alpha(\vec r,t)=\rme^{t/\alpha_0}f^\alpha(\varphi,r).
\]
The normalization condition for the reduced field characteristic function,
namely condition a) in Appendix~\ref{app1}, $w_f(0,t)=1$,
becomes
\[
f^1(\varphi,r)\big|_{\vec r=0}=\rme^{-t/\alpha_0}.
\]
This condition cannot be satisfied for an arbitrary $t$, because the left-hand side
does not depend on time.

Second, let $\alpha_0=0$. Then the invariant solution takes the form
\[
u^\alpha(\vec r,t)=\rme^{\varphi/\beta_0}f^\alpha(t,r).
\]
Using $s \pm \rmi k = r\rme^{ \pm \rmi\varphi}$,
the first component may be written as
\[
u^1
=
\left(
{s+\rmi k\over s-\rmi k}
\right)^{1/(2\rmi\beta_0)}
f^1(t,k^2+s^2).
\]
However, as $\vec r\to0$, the angular factor has no unique limit. One might try
to compensate this behavior by introducing radial power through
$f^1(t,r)$, for instance, using
\[
(s^2+k^2)^a=(s+\rmi k)^a(s-\rmi k)^a,
\]
with some exponent $a$. Nevertheless, this cannot remove the angular dependence
in a manner that yields the required limit for the normalization condition. 
Therefore, this case does not lead to an admissible characteristic function.

Finally, consider the case $\alpha_0\beta_0\neq0$. The invariant solution has
the form
\[
u^\alpha(\vec r,t)=\rme^{t/\alpha_0}f^\alpha(z,r),
\]
where
\[
z=
{k-s\tan\left(\frac{\beta_0}{\alpha_0}t\right)
\over
s+k\tan\left(\frac{\beta_0}{\alpha_0}t\right)}.
\]
Variable $z$ also has no well-defined limit at $\vec r\to0$. Thus, either
$f^1$ must be independent of $z$, so that $f^1=f^1(r)$, or it must depend on a
regularized combination such as $zr^a$, with $a>0$, so that $zr^a\to0$ as
$r\to0$. However, in both cases, the value of $f^1$ at the origin cannot
compensate for the prefactor $\rme^{t/\alpha_0}$ for arbitrary $t$. Consequently,
the normalization condition cannot be satisfied again.

We conclude that the invariant solutions associated with $\Theta_3$ are not
compatible with the boundary conditions imposed by the characteristic-function
description. Therefore, $\Theta_3$ does not yield physically admissible
invariant solutions for the JCM in the present framework.

\section{\label{CL}Conservation laws}

A conservation law associated with a system of differential equations
(\ref{PDEsystem}) is represented by a conserved current as follows:
\begin{equation}\label{Acons}
\vec A=(A^1,\ldots,A^n),
\end{equation}
whose divergence vanishes on the solution manifold of the system. In terms of the 
total derivatives, this condition can be written as:
\begin{equation}
\left.\nabla\cdot\vec A\right|_{\mathcal F=0}
=
\left.
\left(
D_1A^1+\cdots+D_nA^n
\right)
\right|_{\mathcal F=0}
=0,
\end{equation}
where \(D_j\) is the total derivative with respect to \(x^j\), as defined in
Eq.~(\ref{TOTAL_D}). In general, the components \(A^i\) may depend on the
independent variables, dependent variables, and a finite number of their
derivatives,
\[
A^i=A^i(\vec x,\vec u,\vec u^{(1)},\ldots,\vec u^{(s)}).
\]

Determining conserved currents directly is, in general, a nontrivial problem.
However, the methodology introduced by Vinogradov
~\cite{Vinogradov:1984aa,Vinogradov1989_1,Vinogradov1999} provides a systematic
way to associate local conservation laws with their generating functions
\begin{equation}\label{genfuncons}
\vec\psi=(\psi^1,\ldots,\psi^m)  .
\end{equation}
For a system of PDEs called \(l\)-normal system
(see Chapter~5 of Ref.~\onlinecite{Vinogradov1999}), the generating functions are uniquely defined by  conservation laws. Conservation laws that differ by null divergence correspond to the same generating function. The generating function $\vec{\psi}$ is the solution of the following equation (see also Ref. \onlinecite{Anco:1997aa}) 
\begin{equation}\label{sist_func_gen}
\left.
l^\star_{\vec F}\vec\psi
\right|_{\mathcal F=0}
=0.
\end{equation}
The formal adjoint \(^{\star}\) of a scalar differential operator is defined by
\begin{equation}
\left(
\sum_{\varsigma}\alpha_\varsigma D^{(\varsigma)}
\right)^\star
=
\sum_{\varsigma}
(-1)^{|\varsigma|}
D^{(\varsigma)}\circ \alpha_\varsigma .
\end{equation}
For the matrix differential operator \(l_{\vec F}\), the adjoint is obtained by
taking the formal adjoint of each entry and transposing the matrix:
\begin{equation}
l^\star_{\vec F} = \left(\left[l_{\vec F}\right]^\star_{\beta\alpha}\right), \quad {\alpha,\beta=1,\dots,m}.
\end{equation}

A solution \(\vec\psi\) of Eq.~(\ref{sist_func_gen}), sometimes called
{\it adjoint} symmetry or {\it co}-symmetry, corresponds to a conservation law if and only
if there exists a differential operator \(\mathcal T\) such that
\begin{equation}\label{criterio}
l_{\vec\psi}
+
\left.\square\right|_{\mathcal F}^{\star}
=
\left.\mathcal T\right|_{\mathcal F}\circ l_{\vec F},
\qquad
\mathcal T^\star=\mathcal T,
\end{equation}
where $ \square\vec F=l^\star_{\vec F}\vec\psi$, and \(l_{\vec\psi}\) denotes the linearization of \(\vec\psi\). Recall that
a system \(\vec F=0\) is an Euler--Lagrange system if and only if
\[
l_{\vec F}=l^\star_{\vec F}.
\]
If \(\vec\psi\) satisfies criterion (\ref{criterio}), then the associated
conserved current can be obtained from the relation
\begin{equation}\label{CLaw}
\nabla\cdot\vec A = S_\alpha F^\alpha,
\end{equation}
where summation over \(\alpha\) is understood. The operators \(S_\alpha\) are
defined by the condition
\begin{equation}\label{S}
\psi^\alpha
=
\left.
S_\alpha^\star 1
\right|_{\mathcal F=0}.
\end{equation}

We now apply this construction to the JCM differential system
$\vec F=(F^1,F^2,F^3,F^4) =0$ given in Eq.~(\ref{eqs}). The system of equations under consideration is a determined regular system of equations of the evolutionary type (i.e., the derivatives with respect to time are expressed explicitly through derivatives with respect to the remaining variables); therefore, it is $l$-normal. Since the linearization operator
(\ref{l_F_JM}) is written in the variable \(R=r^2/4\), we first work with the
current $\vec A=(A^t,A^R,A^\varphi)$. Thus,
\begin{equation}\label{div}
D_tA^t+D_RA^R+D_\varphi A^\varphi
=
S_\alpha F^\alpha .
\end{equation}

The formal adjoint of the linearization operator (\ref{l_F_JM}) is
\begin{equation}
l^\star_{\vec F}
=
\begin{pmatrix}
-D_t-D_\varphi & 4\vk R & 0 & 0\\
-\vk & -D_t-D_\varphi & -\delta & \vk D_R\\
0 & \delta & -D_t-D_\varphi & \frac{\vk}{2R}D_\varphi\\
0 & 4\vk(1+RD_R) & 2\vk D_\varphi & -D_t-D_\varphi
\end{pmatrix}.
\end{equation}
Since
\[
l_{\vec F}\neq l^\star_{\vec F},
\]
system (\ref{eqs}) is not an Euler--Lagrange system; therefore, the conservation laws cannot be obtained directly from Noether's theorem.
Nevertheless, operators \(l_{\vec F}\) and \(l^\star_{\vec F}\) are related as follows (so called conformally self-adjoint operators):
\[
l^\star_{\vec F}\circ E_R=l_{\vec F},
\qquad
E_R=
\operatorname{diag}
\left(
1,(4R)^{-1},(4R)^{-1},1
\right).
\]
This relation provides a direct connection between symmetry generating
functions and co-symmetries. If \(\vec\phi\) is a generating
function of symmetry that satisfies Eq.~(\ref{l_F_phi}), then
\[
\vec\psi=E_R\vec\phi
\]
satisfies the adjoint equation (\ref{sist_func_gen}). Equivalently,
\[
\vec\phi=E_R^{-1}\vec\psi.
\]
\AY{It should be noted that the conserved currents have meaning if all components $u^\alpha$ of the characteristic functions and the conserved currents themselves are defined as $\vec{r} \to 0$. Although conservation laws hold for all formally defined solutions, in some cases below we will consider conservation laws for the first invariant solution in an attempt to derive constants of motion and provide a physical interpretation for them.}

\subsection{Zero order generating functions}
The zero-order symmetry generating functions are
\[
\vec{\phi}_3=\vec u(\vec r,t),
\qquad
\vec{\phi}_4=\vec{\tilde u}(\vec r,t),
\]
where $\vec{\tilde u}$ is an arbitrary solution of linear system
(\ref{eqs}). From the relationship between symmetries and co-symmetries discussed
above, the corresponding co-symmetries are
\[
\vec{\psi}_3=E_R\vec u,
\qquad
\vec{\psi}_4=E_R\vec{\tilde u}.
\]
In particular,
\begin{equation}
\vec{\psi}_3
=
\left(
u^1,\frac{u^2}{4R},\frac{u^3}{4R},u^4
\right)  .
\end{equation}
For this co-symmetry, the operator $\square$ is given by
$\square=-E_R$, and therefore it is self-adjoint,
$\square^\star=\square$. Moreover, $l_{\vec{\psi}_3}=E_R$, such that
\[
l_{\vec{\psi}_3}+\square^\star=\mathcal O,
\]
where $\mathcal O$ denotes the zero operator. Hence, criterion
(\ref{criterio}) is satisfied with $\mathcal T=\mathcal O$, and
$\vec{\psi}_3$ is the generating function of the conservation law.

From Eq.~(\ref{S}), we have
\[
\vec S=
\left(
u^1,\frac{u^2}{4R},\frac{u^3}{4R},u^4
\right).
\]
Integrating Eq.~(\ref{CLaw}), we obtain the conserved current
\begin{eqnarray}
A^t
&=&
\frac{(u^1)^2+(u^4)^2}{2}
+
\frac{(u^2)^2+(u^3)^2}{8R},
\\
A^R
&=&
-\vk u^2u^4,
\\
A^\varphi
&=&
A^t-\frac{\vk}{2R}u^3u^4.
\end{eqnarray}
By transforming back to the Cartesian Fourier variables $(s,k)$ and using the
inverse transformations (\ref{rotpt}) and (\ref{wfx-yz}), the current
can be written as
\begin{eqnarray}
A^t
&=&
w_{ee}^2+w_{gg}^2-2w_{eg}w_{ge},
\\
A^s
&=&
2\rmi\vk\, w_z w_x-kA^t,
\\
A^k
&=&
2\rmi\vk\, w_z w_y+sA^t .
\end{eqnarray}

\AY{The conservation law can be written in integral form. Consider a disk of radius $r_0 > 0$ centered at $(0,0)$: $C_{r_0} := \{(k,s)\in\mathbb{R}^2: k^2 + s^2 \le r_0^2\}$. Integrating $\operatorname{div} \vec{A} = 0$ over $C_{r_0}$ and applying Green's theorem yields:
\bea
\iint\limits_{C_{r_0}}{D_t A^t dkds} = - \iint\limits_{C_{r_0}}{\left(D_k A^k + D_s A^s\right) dkds} \nonumber\\
= - \oint\limits_{\partial C_{r_0}}\left.\left(A^k, A^s\right)\right|_{r = r_0}\cdot(n_k, n_s) dl, \nonumber
\eea
where $(n_k, n_s)$ is the outward unit normal vector to the boundary $\partial C_{r_0}$ of the region $C_{r_0}$. In polar coordinates, this expression takes the form
\bea
D_t\int\limits_{0}^{r_0}r dr \int\limits_{0}^{2\pi}d\varphi \left[ (u^1)^2 + (u^4)^2 + \left({u^2/ r}\right)^2
+\left({u^3/ r}\right)^2 \right] \nonumber \\
= 4\vk \int\limits_{0}^{2\pi}u^2(r_0,\varphi,t) u^4 (r_0,\varphi,t) d\varphi. \nonumber
\eea
If $\lim_{r\to r_0}u^2(r,\varphi,t) u^4 (r,\varphi,t) = 0$, then the right-hand side equals zero, and a constant  of motion is obtained. For example, if $\lim_{r\to \infty}u^2(r,\varphi,t) u^4 (r,\varphi,t) = 0$, then the following integral does not depend on time:
\be
\iint\limits_{\mathbb{R}^2} \left[w_{ee}^2+w_{gg}^2-2w_{eg}w_{ge}\right] dk ds = {\rm const}. \nonumber
\ee
 }

\AY{
In the context of quantum composite systems, the integral form of a conservation law may not always be straightforward to interpret. In the following, we consider the evaluation of local conservation 
laws at the Fourier-space origin and exploit the characteristic-function correspondence
to turn a local Fourier-space continuity equation into a balance equation for 
observable quantities in the reduced atomic state. The components of the conservative current, calculated at the origin, yield:} 
\begin{eqnarray}
A^t\big|_{\vec r=0}
&=&
P_a(t)-f_a^2(t),
\\
A^s\big|_{\vec r=0}
&=&
2\rmi\vk\langle\sigma_x\rangle\langle\sigma_z\rangle,
\\
A^k\big|_{\vec r=0}
&=&
2\rmi\vk\langle\sigma_y\rangle\langle\sigma_z\rangle,
\end{eqnarray}
where
\begin{equation}
P_a(t)=\varrho_{ee}^2(t)+\varrho_{gg}^2(t)+2|\varrho_{eg}(t)|^2
\end{equation}
is the purity of the reduced atomic density matrix, and
\begin{equation}
f_a(t)=2|\varrho_{eg}(t)|
\end{equation}
\AY{is the normalized atomic coherence function of a maximally coherent superposition of the atomic states.} Thus, the density component of the conserved
current is not the atomic purity alone, but the combination
\AY{
\begin{eqnarray}
P_a(t)-f_a^2(t) &=&\varrho_{ee}^2(t)+\varrho_{gg}^2(t)-2|\varrho_{eg}(t)|^2 \\\nonumber
&=& \langle \sigma_z(t)\rangle^2 + {1\over 2} I^2_{a|f}(t),
\end{eqnarray}
where $I_{a|f}(t) = \sqrt{2 (1- P_a(t)) }$ is the pure state I-concurrence of the 
atom--field bipartite system \cite{RuBuCaHiMi01} which measures the entanglement 
between the atomic and field subsystems through the degree of local mixedness of 
the atomic reduced state; in fact the quantity $1-P_a(t)$ describes 
the linear entropy generated by entanglement with the field. In this regard, this expression provides a 
balance equation involving atomic inversion and an 
entanglement-related contribution}.

The conservation equation
\[
D_tA^t+D_sA^s+D_kA^k=0
\]
then gives, at the origin,
\begin{eqnarray}
{\rmd\over\rmd t}
\left(P_a(t)-f_a^2(t)\right)
&=&
-2\rmi\vk
\left.
\left[
\partial_s(w_z w_x)+\partial_k(w_z w_y)
\right]
\right|_{\vec r=0}
\nonumber\\
&&
-
\left.
\left(\partial_\varphi A^t\right)
\right|_{\vec r=0}.
\label{Purity_flux1}
\end{eqnarray}
This equation shows that in the absence of atom--field coupling
($\vk=0$), and for regular solutions for which the angular contribution at the
origin vanishes, the quantity $P_a-f_a^2$ is conserved. When coupling is
present, its rate of change is controlled by the coupling strength and 
derivatives of the products of the characteristic-function components. 
\AY{These terms contain joint atom-field moments generated by 
the interaction and are therefore sensitive to the development of atom-field correlations.}
By using the relations in Eq.~(\ref{xpn}), Eq.~(\ref{Purity_flux1}) can be written
as
\begin{eqnarray}
{\rmd\over\rmd t}
\left(P_a(t)-f_a^2(t)\right)
&=&
2\vk
\left(
\langle\sigma_z\rangle\langle \hat p_x^1\rangle
+
\langle\sigma_z\rangle\langle \hat x_y^1\rangle
\right.
\nonumber\\
&&
\left.
+
\langle\sigma_x\rangle\langle \hat p_z^1\rangle
+
\langle\sigma_y\rangle\langle \hat x_z^1\rangle
\right)
-
\left.
\left(\partial_\varphi A^t\right)
\right|_{\vec r=0}.\nonumber\\\label{Purity_flux2}
\end{eqnarray}
In this form, the balance equation makes explicit that the variation of
$P_a-f_a^2$ is driven by atom-field coupling and by field moments weighted by
atomic expectation values. Because for a closed bipartite pure state, changes in
the local purity are directly related to the generation of atom-field
entanglement, Eqs.~(\ref{Purity_flux1}) and (\ref{Purity_flux2}) \AY{provide a local Fourier-space balance relation containing an entanglement-related 
contribution in the globally pure case. For mixed global states, $1-P_a$ quantifies 
local mixedness and does not by itself determine the atom--field entanglement.}

\begin{figure}[htbp]
\begin{center}
\includegraphics[width=4.5cm]{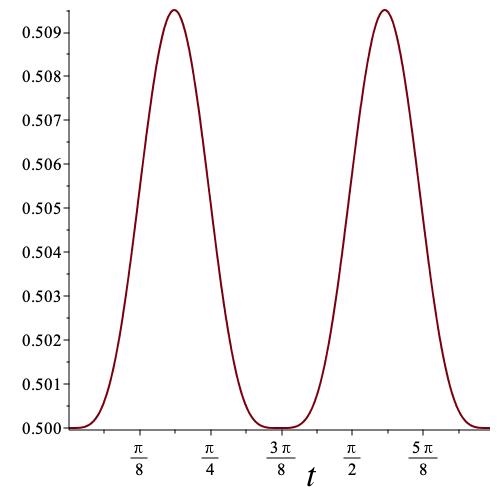}
\caption{Graph of function $P_a(t)$ for $n = 6$, $a = \delta = g = 1$, and $t\in[0, 4\pi/\lambda_6]$.}
\label{fig:2}
\end{center}
\end{figure}

For the invariant solutions derived in Secs.~\ref{subsec:Inv1} and
\ref{subsec:Inv2}, the angular contribution satisfies
\[
\left.
\left(\partial_\varphi A^t\right)
\right|_{\vec r=0}=0.
\]
In particular, for the decoupled-state solutions
(\ref{sol112})--(\ref{sol141}), one obtains
\begin{equation}
\partial_t
\left[
(u^1_n)^2+(u^4_n)^2+\left({u^2_n\over r}\right)^2
+\left({u^3_n\over r}\right)^2
\right]
=
4\vk\,{\partial_r(u^2_nu^4_n)\over r},
\label{CL_0}
\end{equation}
and $f_a(t)\equiv0$. Therefore,
\begin{equation}
2P_a(t)
=
\left[(u^1_n)^2+(u^4_n)^2\right]_{\vec r=0}
=
\frac{4g^4}{\lambda_n^4}
\left(\cos\lambda_n t-1\right)^2+1.
\end{equation}
This explicitly illustrates \AY{(see Fig. \ref{fig:2})} how the local atomic purity oscillates under an 
atom--field interaction: \AY{the balance relation for the first invariant solution, which in this case reduces to the atomic purity alone. The figure shows an initial purity of one-half. Since this solution corresponds to a decoupled atom-field initial configuration (\ie, the initial condition factorizes as described in Eq. (\ref{IN_CON_W})), this indicates that both the atom and the field are initially in mixed states. As time evolves, the increase in atomic purity reflects the buildup of coherences, which correlate with the atomic inversion. This behavior also signals the generation of entanglement between the atom and the field.}

The second zero-order co-symmetry,
\[
\vec{\psi}_4
=
\left(
\tilde u^1,
\frac{\tilde u^2}{4R},
\frac{\tilde u^3}{4R},
\tilde u^4
\right) ,
\]
also generates conservation laws. In this case
$l_{\vec{\psi}_4}=\mathcal O=\square^\star|_{\mathcal F}$, and the associated
current is
\begin{eqnarray}
A^t
&=&
\tilde u^1u^1
+
{\tilde u^2\over 4R}u^2
+
{\tilde u^3\over 4R}u^3
+
\tilde u^4u^4,
\nonumber\\
A^R
&=&
-\vk\left(\tilde u^2u^4+\tilde u^4u^2\right),
\\
A^\varphi
&=&
A^t
-
{\vk\over 2R}
\left(\tilde u^3u^4+\tilde u^4u^3\right).
\nonumber
\end{eqnarray}
This type of conservation law is common in linear systems of differential
equations, where the superposition principle naturally produces conservation
laws involving arbitrary solutions of the adjoint system
(see Example~3.3, Chapter~5 of Ref.~\onlinecite{Vinogradov1999}).

A particularly important consequence is that the usual constant of motion of
the JCM can be recovered from this construction. Consider Eq.~(\ref{CLaw}) with
\[
S_1
=
-D_R-RD_{RR}-{1\over 4R}D_{\varphi\varphi},
\qquad
S_2=S_3=0,
\qquad
S_4=1.
\]
Using Eq.~(\ref{S}), this choice corresponds to the generating function
\[
\vec{\psi}_4=(0,0,0,1) .
\]
Therefore, there exists a current $\vec A$ such that
\[
S_1F^1+F^4=\operatorname{div}\vec A.
\]
In Cartesian variables $(k,s)$, operator $S_1$ is the negative
Laplacian in the Fourier phase space,
\[
S_1=-\Delta_{ks}=-(\partial_{ss}+\partial_{kk}).
\]
Thus,
\begin{equation}\label{OO}
S_1F^1+F^4
=
\partial_tu^4-\Delta_{ks}(\partial_tu^1)
+
\vk R\partial_{RR}u^2
+
O(\varphi)
=
0,
\end{equation}
where $O(\varphi)$ denotes terms containing derivatives with respect to
$\varphi$. For the invariant solutions of Sec.~\ref{subsec:Inv1}, the
$\varphi$-dependent terms vanish. Moreover, from Eq.~(\ref{Y}),
\(
\partial_{RR}u^2=\left(1-{\ell\over R}\right)u^2
\), $u^2|_{R=0} = 0$,
and the radial behavior of the admissible solutions implies
$R\partial_{RR}u^2\to0$, as $R\to0$.
Therefore, evaluation of Eq.~(\ref{OO}) at $\vec r=0$ yields 
\begin{equation}
\left.
\partial_t
\left(
w_z-\partial_{ss}w_f-\partial_{kk}w_f
\right)
\right|_{\vec r=0}
=0.
\end{equation}
Using the characteristic-function identities
\begin{equation}\label{xp}
\langle \hat x^2(t)\rangle
=
-\left.\partial_{kk}w_f(\vec r,t)\right|_{\vec r=0},
\
\langle \hat p^2(t)\rangle
=
-\left.\partial_{ss}w_f(\vec r,t)\right|_{\vec r=0},
\end{equation}
together with
\[
\langle\sigma_z(t)\rangle=w_z(\vec r,t)\big|_{\vec r=0},
\]
we recover the conserved quantity
\begin{equation}
\langle\hat N\rangle
=
\langle\sigma_z(t)\rangle
+
\langle\hat x^2(t)\rangle
+
\langle\hat p^2(t)\rangle
=
\mathrm{const.}
\label{CONSTANT}
\end{equation}
which corresponds, up to the normalization conventions used for the quadrature
variables, to the expectation value of the total number of excitations in the
JCM.  \AY{This result makes an important validation of the conservation-law construction, since one of the local conservation laws in Fourier-space can be explicitly identified with  a known quantum-mechanical constant of motion.}

\subsection{Generating functions of non-zero order}
We now consider the conservation laws generated by co-symmetries that depend on the first and higher derivatives of the dependent variables. Starting from the symmetry
generating functions $\vec\phi_1,\;\vec\phi_2,\; \vec\phi_5$,
given in Eqs.~(\ref{gen_function_phi1}), (\ref{gen_function_phi2}), and
(\ref{gen_function_phi5}), we obtain the corresponding adjoint symmetries
through the relation
\[
\vec\psi_{1,2,5}=E_R\vec\phi_{1,2,5}.
\]
In particular,
\begin{eqnarray}\nonumber
\vec{\psi}_5
\!\!\!\!&=&\!\!\!\!
\left(
{\delta\over\vk}u^4
\!+\!
{\partial_\varphi u^2\over 2R}
\!-\!
\partial_R u^3,
\,
-{\partial_\varphi u^1\over 2R},
\,
-u^4\!+\!\partial_R u^1,
\,
{\delta\over\vk}u^1\!-\!u^3
\right)  .\\
\end{eqnarray}
Among these first-order adjoint symmetries, only $\vec\psi_5$ satisfies 
criterion (\ref{criterio}) and therefore, defines a conservation law. In this
case,
\begin{eqnarray}
l_{\vec{\psi}_5}
&=&
-\square
=
-\square^\star
\nonumber\\
&=&
\begin{pmatrix}
0& (2R)^{-1} D_\varphi & -D_R & \delta/\vk\\
-(2R)^{-1}D_\varphi &0 & 0 & 0\\
D_R & 0 & 0 & -1\\
\delta/\vk & 0& -1 & 0
\end{pmatrix}.
\end{eqnarray}
The associated conserved current, written in terms of
$\vec w=(w_f,w_x,w_y,w_z) $ and the Cartesian Fourier variables $(s,k)$, has
components:
\begin{eqnarray}
A^t
&=&
w_f
\left[
{\delta\over\vk}w_z
+
2\rmi\left(\partial_s w_y-\partial_k w_x\right)
\right]
+
\rmi w_z(sw_y-kw_x),
\nonumber\\
A^s
&=&
-\vk
\left[
k\left(w_f^2+w_z^2-w_x^2+w_y^2\right)
+
2s w_xw_y
\right]
-kA^t,
\\
A^k
&=&
\vk
\left[
s\left(w_f^2+w_z^2+w_x^2-w_y^2\right)
+
2k w_xw_y
\right]
+sA^t.
\end{eqnarray}
Thus, the conservation law is
\[
D_tA^t+D_sA^s+D_kA^k=0.
\]
Equivalently, in polar variables, this equation can be written as
\begin{eqnarray}
\hspace{-0.5cm}D_t A^t + D_\varphi \left(A^t + \vk (w_f^2 + w_z^2)\right) &=& \nonumber \\ 
&&\hspace{-3.5cm}-\vk(r \sin2\varphi D_r  + \cos 2\varphi D_\varphi)(w_x^2 - w_y^2) \nonumber \\ 
&&\hspace{-2.8cm}+ 2\vk (r\cos2\varphi D_r - \sin 2\varphi D_\varphi)(w_x w_y).
\end{eqnarray}

For the invariant solutions (\ref{sol112})--(\ref{sol141}) associated with
$\Theta_1$, one obtains $A^t=-u_n^3u_n^4$,
and the conservation law reduces to
\begin{equation}
\partial_t(u_n^3u_n^4)
=
g\sqrt{2}\,{1\over r}\partial_r(u_n^2u_n^3).
\end{equation}

Higher-order conservation laws can be constructed by applying recursion
operators $\mathrm{D}_t$ and $\mathrm{D}_\varphi$ to the co-symmetries. At the
second order, the following co-symmetries define conservation laws:
\begin{eqnarray}
\vec{\psi}_1^{(t)}
&:=&
\mathrm{D}_t\vec{\psi}_1
=
\left(
\partial_{tt}^2u^1,
{\partial_{tt}^2u^2\over 4R},
{\partial_{tt}^2u^3\over 4R},
\partial_{tt}^2u^4
\right) ,
\\
\vec{\psi}_1^{(\varphi)}
&:=&
\mathrm{D}_\varphi\vec{\psi}_1
=
\left(
\partial_{t\varphi}^2u^1,
{\partial_{t\varphi}^2u^2\over 4R},
{\partial_{t\varphi}^2u^3\over 4R},
\partial_{t\varphi}^2u^4
\right) ,
\\\nonumber
\vec{\psi}_2^{(\varphi)}
&:=&
\mathrm{D}_\varphi\vec{\psi}_2
=
\left(
\partial_{\varphi\varphi}^2u^1,
{\partial_{\varphi\varphi}^2u^2\over 4R},
{\partial_{\varphi\varphi}^2u^3\over 4R},
\partial_{\varphi\varphi}^2u^4
\right) .\\
\end{eqnarray}
The co-symmetries obtained from $\mathrm{D}\vec\psi_5$ ($\mathrm{D}$ denotes  either $\mathrm{D}_t$ or $\mathrm{D}_\varphi$) do not satisfy 
criterion (\ref{criterio}) at this order.

The corresponding conserved currents have the same quadratic structure as 
zero-order current. For $\vec{\psi}_1^{(t)}$, one obtains
\begin{eqnarray}
A^t
&=&
{(\partial_tu^1)^2+(\partial_tu^4)^2\over 2}
+
{(\partial_tu^2)^2+(\partial_tu^3)^2\over 8R},
\\
A^R
&=&
-\vk\,\partial_tu^2\,\partial_tu^4,
\\
A^\varphi
&=&
A^t
-
{\vk\over 2R}\partial_tu^3\,\partial_tu^4.
\end{eqnarray}
For $\vec{\psi}_2^{(\varphi)}$, one obtains
\begin{eqnarray}
A^t
&=&
{(\partial_\varphi u^1)^2+(\partial_\varphi u^4)^2\over 2}
+
{(\partial_\varphi u^2)^2+(\partial_\varphi u^3)^2\over 8R},
\\
A^R
&=&
-\vk\,\partial_\varphi u^2\,\partial_\varphi u^4,
\\
A^\varphi
&=&
A^t
-
{\vk\over 2R}\partial_\varphi u^3\,\partial_\varphi u^4.
\end{eqnarray}
Finally, for $-2\vec{\psi}_1^{(\varphi)}$, the conserved current is
\begin{eqnarray}\nonumber
A^t
&=&
\partial_tu^1\,\partial_\varphi u^1
+
\partial_tu^4\,\partial_\varphi u^4
+
{\partial_tu^2\,\partial_\varphi u^2
+
\partial_tu^3\,\partial_\varphi u^3
\over 4R},\\
&& \\ 
A^R
&=&
-\vk
\left(
\partial_tu^2\,\partial_\varphi u^4
+
\partial_\varphi u^2\,\partial_tu^4
\right),
\\
A^\varphi
&=&
A^t
-
{\vk\over 2R}
\left(
\partial_tu^3\,\partial_\varphi u^4
+
\partial_\varphi u^3\,\partial_tu^4
\right).
\end{eqnarray}

For invariant solutions (\ref{sol112})--(\ref{sol141}), only the first of
these second-order fluxes is nontrivial. This yields a conservation law analogous
to Eq.~(\ref{CL_0}):
\begin{eqnarray}
\partial_t
\left[
(\partial_tu_n^1)^2
+
(\partial_tu_n^4)^2
+
\left({\partial_tu_n^2\over r}\right)^2
+
\left({\partial_tu_n^3\over r}\right)^2
\right]
=\nonumber \\
4\vk\,
{\partial_r\left(\partial_tu_n^2\,\partial_tu_n^4\right)\over r}.
\end{eqnarray}

At the third order, applying $\mathrm{D}^2$ to $\vec\psi_1$ and $\vec\psi_2$ does
not generate new conservation laws, whereas applying $\mathrm{D}^2$ to
$\vec\psi_5$ does. For instance, the co-symmetry
$\mathrm{D}_t^2\vec\psi_5$ generates the current
\begin{eqnarray}
A^t
&=&
{\delta\over\vk}\partial_tu^1\,\partial_tu^4
-
\partial_tu^3\,\partial_tu^4
+
\partial_{tR}u^1\,\partial_tu^3
+
{\partial_tu^1\,\partial_{t\varphi}u^2\over 2R},
\\
A^R
&=&
-\partial_tu^1\,\partial_{t\varphi}u^3
-
\partial_tu^1\,\partial_{tt}u^3,
\\
A^\varphi
&=&
A^t
+
\vk(\partial_tu^1)^2
+
\vk(\partial_tu^4)^2
+
{\vk\over 4R}
\left[
(\partial_tu^2)^2+(\partial_tu^3)^2
\right]
\nonumber\\
&&
-
{\partial_{tt}u^1\,\partial_tu^2
+
\partial_{t\varphi}u^1\,\partial_tu^2
\over 2R}.
\end{eqnarray}

This pattern continues to higher order. In particular, repeated applications
of recursion operators $\mathrm{D}_t$ and $\mathrm{D}_\varphi$ generate an
infinite hierarchy of conservation laws. More precisely,
\[
\mathrm{D}^{2m+1}\vec\psi_{1,2},
\qquad
\mathrm{D}^{2m}\vec\psi_5,
\qquad
m=0,1,\ldots,
\]
provide generating functions of conservation laws. 


\section{\label{summ}Summary}

In this work, we applied the Lie symmetry theory to the system of partial
differential equations describing the Jaynes--Cummings model (JCM) in the
characteristic-function representation. The purpose of this analysis was to
characterize the model from a symmetry-based perspective, identify its admitted
invariant solutions, and investigate the conservation laws encoded in its
differential formulation. The physical admissibility of the resulting
solutions is determined by the quantum-mechanical conditions imposed on the
field characteristic function and the reduced atomic density matrix, which
act as the boundary and regularity conditions for the PDE system.

The symmetry analysis identifies two relevant classes of invariant solutions that are 
compatible with these requirements. The first class recovers familiar
dressed-state dynamics from the invariant reduction of differential
equations, thereby providing a different characterization of the standard JCM
solution. The second class has a radial dependence expressed in terms of Heun polynomials and describes generic coupled atom--field
configurations. \AY{Rather than lying outside the conventional 
Fock-space description of the Jaynes-Cummings model, the structure 
of this invariant solution indicates that its Heun-polynomial contribution 
can be interpreted in terms of field coherences connecting photon-number states 
separated by $2m$ excitations.}  Other mathematically admissible invariant forms were found to be
incompatible with the normalization and regularity conditions required for a
physical characteristic function.

The differential system also possesses a nontrivial conservation-law
structure. Using the methodology developed by Vinogradov, we recovered the
well-known conservation of the total number of excitations directly from the
PDE formulation. In addition, we derived conserved currents involving atomic
populations, coherence, reduced-state purity, and moments of field
characteristic functions. Of particular interest is the balance equation for
the quantity $P_a-f_a^2$, \AY{which for globally pure states, 
relates it to the atom-field entanglement; however, the 
resulting balance equation does not constitute a conservation law 
for entanglement itself. On the other hand, its} rate of change is determined by the
atom--field coupling and by combinations of atomic expectation values with
field-related characteristic-function moments.

These results should not be interpreted as establishing a conservation law for the 
entanglement itself. Rather, they show that the differential equations contain
conserved currents involving quantities whose evolution is linked to the
generation and redistribution of atom--field correlations. For globally pure
states, variations of the reduced atomic purity are directly associated with
variations of the bipartite atom--field entanglement. Therefore, the derived balance laws provide a possible mathematical description of how purity, coherence,
and correlations are redistributed during the exchange of excitations between
the two subsystems.

The JCM differential system also exhibits a rich mathematical structure. In
addition to point symmetries, it admits \AY{the genuinely first order generalized symmetry}.
The existence of recursion operators allows generalized symmetries of
arbitrarily high order to be constructed. This is a remarkable property, since relatively few systems admit higher-order symmetries. Although the system is not of the 
Euler--Lagrange type and Noether's theorem cannot be applied directly, a
mapping between symmetries and co-symmetries can be established. This mapping
facilitates the construction of generating functions for conservation laws and
leads to an infinite hierarchy of conserved currents.

\AY{The theorems proved by Vinogradov resolved the problem of the triviality of given conserved currents for $\ell$-normal systems of equations. Since the considered system is  determined system of evolutionary type (therefore, is $\ell$-normal) and the application of recursion operators yields non-trivial generating functions of arbitrary order, the corresponding conservation laws are non-trivial.

To the best of the authors' knowledge, the conservation laws for the system  under consideration are new. It is difficult to determine their physical meaning -- particularly that of the higher-order generating functions. This situation common for many other equations in mathematical physics, where only lower-order conservation laws lend themselves to interpretation, whereas the existence of non-trivial conservation laws of arbitrary order serves as a distinctive property of these equations.

}
Overall, the application of Lie symmetry analysis and Vinogradov's
conservation-law framework provides a complementary perspective on the JCM.
It recovers the known features of the model from the structure of its differential
equations, identifies an additional class of invariant solutions, and reveals
conservation laws that are not immediately apparent from the usual Hamiltonian
treatment. These results suggest that symmetry and conservation-law methods may
also be useful for investigating more complex composite quantum systems whose
dynamics can be formulated as coupled systems of differential equations.

\begin{acknowledgements}
L.M.P. acknowledges a doctoral fellowship from SECIHTI.
The authors used ChatGPT (OpenAI) to assist with English-language editing 
and stylistic revision of the manuscript. The authors reviewed and approved 
the resulting text and took full responsibility for the scientific content.
\end{acknowledgements}

\paragraph*{\bf Data Availability:}
All the data supporting the findings of this study are included in the manuscript.

\paragraph*{\bf Conflict of Interest Declaration:} The authors declare that they have no
affiliations with or involvement in any organization or entity with any financial interest in the subject matter or materials discussed in this manuscript.

\appendix
\section{\label{app1}The characteristic function  \AY{and its physical 
admissibility requirements}}
The characteristic function $w(\vec{r},t)$ (or chord function)
~\cite{Ozo04} is defined as the
Fourier transform of the Wigner function~\cite{Case08}
\begin{eqnarray}\label{Wtransf}
w(\vec{r},t)&=& \iint\rmd p\, \rmd q\; \rme^{\rmi q k+\rmi s p}\, W(q,p,t) \\\nonumber
&=& \int\rmd q\; \rme^{\rmi q k}\, \la q+s/2 |\, \varrho(t)\, | q-s/2\ra \; .
\end{eqnarray}
\AY{From the above definition, the characteristic function can be analogously defined 
as the expectation value of the Weyl operator, \ie:
\begin{equation}
  w(\vec{r},t) = \tr\left[\varrho(t)\, \hat{w}(\vec{r}\,)\right],\qquad \hat{w}(\vec{r}\,)=\rme^{\rmi (k \hat{x} + s \hat{p})},
\end{equation}
where $\hat{x}$ and $\hat{p}$ are the position and momentum operators, respectively.}
This type of transformation maps the quantum mechanical 
phase-space representation $\{q,p\}$, where the quasi-probability
distribution function, or the Wigner function,  ${W}(q,p,  t)$  is defined, 
into its Fourier image $\{k,s\}$. 
Additionally, the application of position and momentum operators to
the density operator can be readily translated into
the characteristic function representation as
(for convenience, higher powers of $\hat x$ and $\hat p$ are included):
\begin{eqnarray}\label{A3}
\h{x}^n \h{p}^m\, \varrho &\mapsto&
   \Big( \frac{s}{2} - \rmi \partial_k\Big)^{n}
   \Big( \frac{-k}{2} - \rmi\partial_s\Big)^{m}w\\
\varrho\, \h{x}^{n}\h{p}^{m} &\mapsto&
   \Big( \frac{-s}{2} - \rmi \partial_k \Big)^n
   \Big( \frac{k}{2} - \rmi\partial_s \Big)^m\; w\\
\h{x}^{n}\,\varrho\, \h{p}^{m} &\mapsto&
   \Big( \frac{s}{2} - \rmi \partial_k\Big)^{n}
   \Big( \frac{k}{2} - \rmi\partial_s\Big)^{m}w\\\label{A6}
\h{p}^{m}\,\varrho\, \h{x}^{n} &\mapsto&
   \Big( \frac{-s}{2} - \rmi \partial_k\Big)^n
   \Big( \frac{-k}{2} - \rmi\partial_s \Big)^m\! w\, .
\end{eqnarray}
This also allows us to obtain explicit expressions for the
$n$-th order moments of the 
products of the position and momentum operators by taking the appropriate
partial derivatives at the origin of the coordinate system:
\begin{eqnarray}\label{xpn}
\frac{\la \h{x}^n\,\h{p}^m \ra + \la \h{p}^m\,\h{x}^n \ra}{2}
 = (-\rmi)^{n+m} \, \partial^n_k\, \partial^{m}_s \, w \big|_{k,s = 0}\; ,
\end{eqnarray}
which also include the cases:
$\la \h{x}^{n}\ra = (-\rmi\, \partial_k)^n\, w \big|_{k,s = 0}$, and $\la
\h{p}^{n}\ra = (-\rmi\,\partial_s)^n\, w \big|_{k,s = 0}$.

For any characteristic function to represent a valid quantum state 
of a density operator $\varrho$, it must 
satisfy some important properties: 
\begin{itemize}
\item[$a)$] the normalization condition 
$w(\vec{r}=0,t) = 1$, 
\item[$b)$] the Hermiticity condition: $w(-\vec{r},t) =w^\ast(\vec{r},t)$, 
\item[$c)$] it has to be continuous and bounded, and 
\item[$d)$] it has to satisfy the so called 
twisted positive definiteness (quantum Bochner condition)~\cite{Brocker95,Quang2016,Dangniam_2015}: 
\AY{
\begin{equation}\label{twisted_pos}
\mathcal{B}_{w}[f]=\int_{\mathbb{R}^4}  \rmd^{2}\vec{r}\;\rmd^{2} \vec{r}\,' f^*(\vec{r}\,)
\rme^{{\rmi\over 2} \sigma(\vec{r},\vec{r}\,')}\, w(\vec{r}-\vec{r}\,',t)f(\vec{r}\,') \geq 0.
\end{equation}
}
where
\begin{equation}
\sigma(\vec{r},\vec{r}\,') = \vec{r} \begin{pmatrix}
0 & 1 \\
-1 & 0
\end{pmatrix}\vec{r}\,',
\end{equation}
\AY{is a symplectic form arising in the condition due to canonical commutation relations, 
and} $f(\vec{r}\,)\in L^2(\mathbb{R}^2)$ is a quadratic integrable test function.
\end{itemize}
\AY{ The quantum Bochner condition is a rigorous mathematical condition that ensures the validity 
any characteristic function as a representation of a quantum state. 
It is a necessary condition for the existence of a density operator 
$\varrho$ corresponding to the characteristic function $w(\vec{r},t)$.

The positivity condition for the density operator $\varrho$ must be satisfied both for the reduced systems and for the complete atom--field system.
In the case of the reduced atomic density operator $\varrho_a$ the positivity 
condition is meet whenever the Bloch vector 
$\vec{b}(t) = (\langle \sigma_x(t) \rangle, \langle \sigma_y(t) \rangle, \langle \sigma_z(t) \rangle)$
satisfies the condition $|\vec{b}(t)| \leq 1$.
Note that for the invariant solutions corresponding to the subalgebras $\Theta_1$ and $\Theta_2$, this condition is satisfied thanks to the appropriate choice of initial conditions for the functions $U^{3,4}(r)$.}

\AY{For the solutions containing the 
Heun polynomials, the validation of the positivity condition can be done 
by writing solution (\ref{eq:full_wf_even}) into its density operator form
as given in Eqs.~(\ref{eq:rho_total_field_diag}) and (\ref{eq:rho_heun}), 
where the diagonal part $\varrho_{\mathrm{diag}}$ is a positive operator,
and the off-diagonal part $\Delta\varrho_{f}(t)$ is a bounded operator. 
In this context, the functional quantum-Bochner problem has therefore been converted
into a matrix positivity problem, for which the positivity of the total density operator 
$\varrho_{f}(t)$ can be verified by checking the positivity of its principal minors or equivalently, 
the positivity of all of its eigenvalues. 

The characteristic function of the composite system is matrix-valued in 
the atomic subspace (see Eq.~(\ref{dmatblocks})), 
and full positivity requires the corresponding matrix-valued quantum Bochner condition, 
rather than the scalar condition for the reduced-field characteristic function 
alone. A generalization of the quantum Bochner condition to the matrix-valued 
case can be derived straightforwardly by defining 
an arbitrary two-component square-integrable test function\cite{Quang2016,Dangniam_2015}:
$\vec{f}(\vec{r}\,) = ( f_e(\vec{r}\,), f_g(\vec{r}\,) ) \in L^{2}(\mathbb{R}^{2})\otimes\mathbb{C}^{2}$. Thus, 
the generalized quantum Bochner condition for the matrix-valued 
characteristic function 
$\mathrm{w}(\vec{r},t)$ takes the form:
\begin{equation}
\begin{aligned}
\mathcal{B}_{\mathrm{w}}[\vec{f}]
=&
\int_{\mathbb{R}^4}\rmd^{2}\vec{r}\;\rmd^{2}\vec{r}\,'
\exp\left[
\frac{\rmi}{2}
\sigma(\vec{r},\vec{r}')
\right]
\\
&\times
\sum_{\mu,\nu=e,g} f_{\mu}^{*}(\vec{r}\,)
w_{\mu\nu}(\vec{r}-\vec{r}\,',t)
f_{\nu}(\vec{r}\,') \geq 0
\end{aligned}
\label{eq:matrix_bochner}
\end{equation}
for every pair of test functions $f_{e}$ and $f_{g}$. In recent studies, the positivity problem for trace-class Weyl operators is explicitly characterized as non-trivial, and numerical as well as discrete approximations are being developed~\cite{Cordero17,Homa_2023}.

The Heun polynomial solutions obtained above
are exact solutions of the JCM characteristic-function PDE satisfying
the necessary normalization, Hermiticity, and regularity
requirements, while their physical admissibility additionally depends on positivity
constraints on the free functions and coefficients. The full physical subset
may in principle be fixed by imposing such constraints, but its complete
characterization requires a dedicated positivity analysis of the block-atom
projected characteristic function that goes beyond the symmetry-based
construction developed here. We regard this explicit characterization of the
positive subset as a natural and important continuation of the present work,
to be presented in a forthcoming study.
}

\section{\label{appJCM} Revisiting the Jaynes-Cummings model}
The standard Jaynes-Cummings model (JCM) describes the interaction between a
two-level atom and a single quantized mode of the electromagnetic field through
the exchange of energy quanta. Its usual analytical treatment relies on the
existence of a constant of motion associated with the total number of
excitations in the composite atom-field system. In a dimensionless formulation
defined with respect to the time scale of the field dynamics, this constant of
motion is represented by the total excitation-number operator
\AY{
\be
\hat N = \frac{1}{2}\sigma_z\otimes \One + \One\otimes \left(\hat n + \frac{\One}{2}\right),
\ee
where 
$\sigma_z$ and $\h{n}$ denote, respectively, the atomic inversion and field number
operator.}

The eigenstates of $\h{N}$ define invariant excitation subspaces of the
atom-field Hilbert space. In particular, each subspace is spanned by bare
states
\[
\{|e,n\rangle, |g,n+1\rangle\}\subset
\mathcal{H}_a\otimes\mathcal{H}_f,
\]
where $|e\rangle$ and $|g\rangle$ denote the excited and ground states of the
two-level atom, while $|n\rangle$ represents a number state of the field mode.

Since the total number of excitations is conserved, operator $\h{N}$
commutes with the Hamiltonian, $[\h{H},\h{N}]=0$. Therefore, the dynamics
decomposes into invariant two-dimensional subspaces, and in the basis
$\{|e,n\rangle,|g,n+1\rangle\}$ the Hamiltonian is represented as
\AY{
\begin{equation}\label{Hn}
  \h{H}\left(\begin{array} {c}
              |g,n+1\rangle \\
              |e,n\rangle
             \end{array} \right)
  =
  \left(\begin{array} {cc}
          -{\Delta\over 2} + n + 1 & g\sqrt{n+1} \\
          g\sqrt{n+1} &  {\Delta\over 2} + n 
        \end{array} \right)
  \left(\begin{array} {c}
          |g,n+1\rangle  \\
          |e,n\rangle
        \end{array} \right).
\end{equation}}

The dressed states, namely the eigenstates of the JCM Hamiltonian in each
excitation subspace, are obtained by diagonalizing this matrix. Thus,
\begin{eqnarray}\label{Heigen}
  \h{H}\left(\begin{array} {c}
              |+\rangle \\
              |-\rangle
             \end{array} \right)
  &=&
  \left(\begin{array} {cc}
          \eps_+ & 0 \\
          0 & \eps_-
        \end{array} \right)
  \left(\begin{array} {c}
          |+\rangle \\
          |-\rangle
        \end{array} \right),
\end{eqnarray}
where the corresponding eigenvalues are \AY{
$\eps_+ = \eps_n + \lambda_n/2$ and
$\eps_- = \eps_n - \lambda_n/2$,} with $\eps_n=n+1/2$. The quantity
\begin{equation}\label{lambn}
  \lambda_n = \sqrt{\delta^2 + 4g^2(n+1)}
\end{equation}
is the generalized Rabi frequency, where $\delta=\Delta-1$ is the detuning.
The resonant regime corresponds to $\delta=0$, for which the atomic transition
frequency coincides with the field-mode frequency.

The dressed states are related to the bare states through the transformation
\begin{equation}\label{vtrans}
\left(\begin{array} {c}
            |+\rangle \\
            |-\rangle
      \end{array} \right)
=
\left(\begin{array} {cc}
        \cos(\theta_n) &  \sin(\theta_n) \\
        -\sin(\theta_n) &  \cos(\theta_n)
      \end{array} \right)
\left(\begin{array} {c}
        |e,n\rangle \\
        |g,n+1\rangle
      \end{array} \right),
\end{equation}
where
\begin{equation}
\theta_n =
\arctan\left(\frac{\lambda_n-\delta}{2g\sqrt{n+1}}\right)
\end{equation}
is the mixing angle between the bare atom-field states. Similarly, 
$\tan 2\theta_n = (2g\sqrt{n+1})/\delta,$ whereas 
\begin{eqnarray} \label{thetan}
\cos\theta_n=\sqrt{\frac{\lambda_n+\delta}{2\lambda_n}}, &\quad&
\sin\theta_n=\sqrt{\frac{\lambda_n-\delta}{2\lambda_n}}.
\end{eqnarray}

Once the dressed states and their eigenvalues are known, the evolution of any
pure state within a fixed excitation subspace can be expressed in this basis.
For an initial state: $|\psi\rangle = c_+|+\rangle + c_-|-\rangle$,
with $c_+,c_-\in\mathbb{C}$ and $|c_+|^2+|c_-|^2=1$, the time evolution is
given by
\AY{\begin{equation}\label{soldress}
  |\tilde{\psi}(t)\rangle =
  c_+ \rme^{-\rmi \lambda_n\, t/2}|+\rangle
  + c_- \rme^{\rmi \lambda_n\, t/2}|-\rangle.
\end{equation}
where $|\tilde{\psi}(t)\rangle=\rme^{\rmi\eps_n t}|\psi(t)\rangle$ has 
been defined to absorb the global phase factor $\rme^{\rmi\eps_n t}$.
} This state can be
written in the bare basis as
\begin{equation}\label{psit}
  |\tilde{\psi}(t)\rangle =
  a_n(t)|e,n\rangle + b_n(t)|g,n+1\rangle,
\end{equation}
where 
\begin{eqnarray}\label{atbt}
  a_n(t) &=&
  c_+\cos\theta_n\,\rme^{-\rmi\lambda_n t/2}
  - c_-\sin\theta_n\,\rme^{\rmi\lambda_n t/2},
  \nonumber \\
  b_n(t) &=&
  c_+\sin\theta_n\,\rme^{-\rmi\lambda_n t/2}
  + c_-\cos\theta_n\,\rme^{\rmi\lambda_n t/2}.
\end{eqnarray}
These amplitudes satisfy the normalization condition
$|a_n(t)|^2+|b_n(t)|^2=1$ for all times.

The dressed-state representation provides a standard analytical route for
solving the JCM in each excitation subspace. However, more general initial conditions,
 may involve superpositions over many photon-number sectors, mixed
states, or field-state representations that are not most conveniently handled
by working directly in the dressed-state basis. In such cases, the system
evolution can be described using the density operator,
\[
\varrho(t)=\h{U}(t)\varrho(0)\h{U}^{\dag}(t),
\]
where $\h{U}(t)=\exp(-\rmi\h{H}t)$ is the evolution operator. Since the
Hamiltonian is block-diagonal in the excitation subspaces, the evolution
operator can be written as
\AY{
\begin{equation}
\h{U}(t)=
\rme^{\rmi\Delta t /2 }|g,0\rangle\langle g,0|
+
\sum_{n=0}^{\infty}\rme^{-\rmi H_n t}\Pi_n,
\end{equation}}
where
\begin{equation}
\Pi_n =
|e,n\rangle\langle e,n|
+
|g,n+1\rangle\langle g,n+1|
\end{equation}
is the projector onto the $n$-th excitation subspace, and $H_n$ denotes the
matrix representation in Eq.~(\ref{Hn}). Thus, the standard construction
requires diagonalization of each excitation block.

Alternatively, the evolution operator may be obtained through direct
matrix-exponentiation methods, yielding a field-number-operator-dependent
representation~\cite{Phoenix88,Klimov09}:
\begin{equation}
  \h{U}(t) =
  \rme^{-\rmi \h{n} t}
  \left(\begin{array} {cc}
    \cos (\h{\lambda} t)
    - \rmi \h{\gamma}\sin ( \h{\lambda} t )
    &
    -\h{\Omega}\sin ( \h{\lambda} t )
    \\
    \rmi \h{\Omega} \sin ( \h{\lambda} t )
    &
    \cos (\h{\lambda} t)
    + \rmi \h{\gamma} \sin ( \h{\lambda} t )
  \end{array} \right),
\end{equation}
where
\AY{
\begin{equation}
\h{\lambda}=\frac{1}{2}\sqrt{\delta^2+4g^2(\h{n}+1)}, \quad
\h{\gamma}=\frac{\delta}{\h{\lambda}}, \quad
\h{\Omega}=\frac{g\sqrt{\h{n}+1}}{\h{\lambda}}.
\end{equation}
}
This representation can be directly applied to the field states expanded in the
number-state basis.

\subsection{Dressed state solution of the JCM in the characteristic function
representation}
The solution of the JCM in the dressed and bare state basis is given by (\ref{soldress}), (\ref{psit}). 
The transformation of this solution into the characteristic function representation is achieved by applying transformation (\ref{Wtransf}) to each component of the density operator. 
The density operator in the bare state basis is given by:
\begin{eqnarray}
  \varrho(t) &=&  |a_n(t)|^2  | e,n \rangle \langle e,n | + |b_n(t)|^2 | g,n+1 \rangle \langle g,n+1 | \\\nonumber
  &+& a_n(t) b_n^*(t) | e,n \rangle \langle g,n+1 | + b_n(t) a_n^*(t) | g,n+1 \rangle \langle e,n |\,,
\end{eqnarray}
thus the components of the density operator can be identified as:
\begin{eqnarray}
  \varrho_{ee}(t) &=& |a_n(t)|^2 | n \rangle \langle n |,\\
  \varrho_{gg}(t) &=& |b_n(t)|^2 | n+1 \rangle \langle n+1 |,\\
  \varrho_{eg}(t) &=& a_n(t) b_n^*(t) | n \rangle \langle n+1 |,\\
  \varrho_{ge}(t) &=& b_n(t) a_n^*(t) | n+1 \rangle \langle n |\,.
\end{eqnarray}
Using the definition of the characteristic function (\ref{Wtransf}) and the position
representation of the number states:
\begin{equation}
  \psi_{n}(x)=\langle x | n \rangle =
  \frac{1}{\sqrt{2^n n! \sqrt{\pi}}} \, H_n\left( x \right)
  \rme^{- x^2/2}\,,
\end{equation}
where $H_n(x)$ are the Hermite polynomials, we obtain the dressed state 
density matrix solution of the JCM as 
\begin{equation}
  \varrho(\vec{r},t) =
  \begin{pmatrix}
    w^{(n)}_{ee}(\vec{r},t) & w^{(n)}_{eg}(\vec{r},t) \\
    w^{(n)}_{ge}(\vec{r},t) & w^{(n)}_{gg}(\vec{r},t)
  \end{pmatrix}\,,
\end{equation}
with the density matrix components described in terms of the
field characteristic function as:
\begin{eqnarray}
  w^{(n)}_{ee}(\vec{r},t\,) &=& |a_n(t)|^2 \, \chi_{n,n}(\vec r),\label{W_ee}\\
  w^{(n)}_{gg}(\vec{r},t\,) &=& |b_n(t)|^2 \, \chi_{n+1, n+1}(\vec r),\\
  w^{(n)}_{eg}(\vec{r},t\,) &=& a_n(t) b_n^*(t) \, \chi_{n, n+1}(\vec r),\\
  w^{(n)}_{ge}(\vec{r},t\,) &=& b_n(t) a_n^*(t) \, \chi_{n+1,n}(\vec r), \label{W_ge}
\end{eqnarray}
where 
\AY{
\begin{eqnarray}\label{chiee}
  \chi_{n,n}(\vec{r}\,) &=& \rme^{-r^2/4}\, L_n(r^2/2) , \\ \label{chigg}
  \chi_{n+1,n+1}(\vec{r}\,) &=& \rme^{-r^2/4}\, L_{n+1}(r^2/2), \\ \label{chieg}
  \chi_{n,n+1}(\vec{r}\,) &=& - \frac{ s - \rmi k}{\sqrt{2(n+1)}} \,
     { \, \rme^{-r^2/4} } \, L_{n}^{(1)}(r^2/2), \\ \label{chige}
 \chi_{n+1,n}(\vec{r}\,) &=&  \frac{ s + \rmi k}{\sqrt{2(n+1)}} \,
     {  \rme^{-r^2/4}}\, L^{(1)}_{n}(r^2/2),
\end{eqnarray}
denotes the characteristic functions transformation of the respective field operators: 
$|n\rangle\langle n|$, $|n+1\rangle\langle n+1|$, $|n\rangle\langle n+1|$, $|n+1\rangle\langle n|$, and $L_n(x)$ $L_n^{(m)}(x)$ are the Laguerre 
and the associated Laguerre polynomials, respectively.} 

\AY{Moreover, any operator of the Fock space $|n\rangle\langle n+d|$, 
for $d\geq0$ and representing coherences between the field 
Fock states $|n\rangle$ and $|n+d\rangle$ is translated
to the characteristic function representation $\chi_{n,n+d}(\vec{r}\,) = 
\langle n+d| \hat w(\vec{r}\,) |n\rangle$ as:
\begin{equation}
\chi_{n,n+d}(\vec{r}\,) =
\sqrt{\frac{n!}{(n+d)!}}\,
\rme^{-\frac{r^{2}}{4}}
\left(
\frac{-s+\rmi k}{\sqrt{2}}
\right)^{d}
L_{n}^{(d)}
\left(
\frac{r^{2}}{2}
\right). \label{eq:fock_element_1}
\end{equation}
Similarly,
\begin{equation}
\chi_{n+d,n}(\vec{r}\,)=
\sqrt{\frac{n!}{(n+d)!}}\,
\rme^{-\frac{r^{2}}{4}}
\left(
\frac{s+\rmi k}{\sqrt{2}}
\right)^{d}
L_{n}^{(d)}
\left(
\frac{r^{2}}{2}
\right).
\label{eq:fock_element_2}
\end{equation}
By taking $d=0$ one immediately recovers the diagonal element 
transformation: 
\[
\chi_{n,n}(\vec{r}\,)=\rme^{-r^{2}/4}
L_{n}\left(
\frac{r^{2}}{2}
\right).
\]
}

Within this representation, the field dynamics can be 
obtained by tracing out the atom degrees of freedom, yielding: 
\AY{
\begin{eqnarray}\nonumber
  w^{(n)}_f(r,t) &=& w^{(n)}_{ee}(r,t) + w^{(n)}_{gg}(r,t)\\\nonumber
  & =& |a_n(t)|^2\chi_{n,n}(\vec{r}\,) \\\label{w_f_CHI} 
  &&\hspace{1cm}+ |b_n(t)|^2\chi_{n+1,n+1}(\vec{r}\,)\,, 
  \end{eqnarray}
}
while the characteristic function description of the 
atom population in this representation, \ie:
$w^{(n)}_z(\vec{r},t) = \tr_a\{\sigma_z \varrho(\vec{r},t)\}$ is given by:
  \AY{
\begin{eqnarray}\nonumber
w^{(n)}_z(r,t) &=& w^{(n)}_{ee}(r,t) - w^{(n)}_{gg}(r,t)\\\nonumber
  & =& |a_n(t)|^2\chi_{n,n}(\vec{r}\,)\\
  &&\hspace{1cm}- |b_n(t)|^2\chi_{n+1,n+1}(\vec{r}\,).
  \end{eqnarray}
  }

By considering 
\begin{eqnarray}\nonumber
a_n(t) &=& c_+ \cos\theta_n\, \rme^{-\rmi t \lambda_n/2} - c_- \sin\theta_n\, \rme^{\rmi t \lambda_n/2},  \\
b_n(t) &=& c_+ \sin\theta_n\, \rme^{-\rmi t \lambda_n/2} + c_- \cos\theta_n\, \rme^{\rmi t \lambda_n/2},
\end{eqnarray}
with $\cos\theta_n$ and $\sin\theta_n$ defined as in (\ref{thetan}), and by taking 
\begin{equation}
c_{+,-} = \sqrt{\frac{1}{2} \mp \frac{\delta}{2\lambda_n}}, 
\end{equation}
we have
\begin{eqnarray}
a_n(t) &=& -\rmi \frac{2g\sqrt{n + 1}}{\lambda_n} \sin\frac{\lambda_n t}{2}, \nonumber \\
b_n(t) &=& \cos\frac{\lambda_n t}{2} +  \rmi \frac{\delta}{\lambda_n} \sin\frac{\lambda_n t}{2}.
\end{eqnarray}
Thus, for example, Eq. (\ref{w_f_CHI}) converts to the following expression
\begin{equation}
w^{(n)}_f(\vec{r},t) = g^2 \frac{1 - \cos\lambda_n t}{\lambda_n^2} Y_n(r) + \rme^{-r^2/4} L_{n+1}(r^2/2),
\end{equation}
\AY{where $Y_n(r)$ as in (\ref{Y_n_alt}).}

After transforming the characteristic-function components
(\ref{W_ee})--(\ref{W_ge}) to the variables \AY{$\vec{u}$ and denoting it as} 
$\vec{u}_\chi(\vec{r},t)$, we find that the resulting solution is related to
the invariant solution (\ref{sol112})--(\ref{sol141}) through the
superposition principle of the linear system. More precisely, the two solutions
coincide after adding to $\vec{u}_\chi$ the stationary solution
\[
\vec{u}_{\mathrm{st}}(\vec{r})
\!:=\!
\left(
-\rme^{-{r^2\over 4}}L_{n+1}(r^2/2)
-\frac{2Y_n(r)}{r^2},
\,
0,
\,
0,
\,
\rme^{-{r^2\over 4}}L_{n+1}(r^2/2)
\right).
\]
Thus,
\[
\vec{u}(\vec{r},t)
=
\vec{u}_\chi(\vec{r},t)
+
\vec{u}_{\mathrm{st}}(\vec{r}),
\]
with the constant chosen as
\[
c=-\frac{g\sqrt{2}}{\lambda_n}.
\]
Using this choice, $\vec{u}(\vec{r},t)$ reproduces the invariant solution
given in Eqs.~(\ref{sol112})--(\ref{sol141}).

\AY{\section{\label{app:C} Proof of the polynomial nature of the Heun functions for the second invariant solution}}
The coefficients $v_k$ in the serie (\ref{H_SERIES}) are determined by the three-term recurrence relation \cite{Fiziev_2010}:
\be
A_k v_k = B_k v_{k - 1} + C_k v_{k - 2}, \qquad v_{-1} = 0, \ v_0 = 1,
\label{recurrence}
\ee
where
\bea
&A_k = 1 + \frac{\beta_h}{k}, \quad C_k = \frac{\alpha_h}{k^2}\left(\frac{\delta_h}{\alpha_h} + \frac{\beta_h + \gamma_h}{2} + k - 1\right), 
\label{ABC1}
\\
&B_k = 1 - \frac{\alpha_h - \beta_h - \gamma_h + 1}{k} + \frac{\eta_h}{k^2} + \frac{\alpha_h - \beta_h - \gamma_h + \beta_h(\gamma_h - \alpha_h)}{2k^2}.
\label{ABC2}
\eea
Let us define 
\be
S_p := B_{p + 1} S_{p - 1} + C_{p + 1} A_p S_{p - 2}, \quad S_{-1} := 1,\ S_0 := B_1, 
\label{S_p}
\ee
for $p\in\mathbb{N}$. Then, from (\ref{recurrence}) we have:
\be
v_1 = \frac{S_0}{A_1}, \ v_2 = \frac{S_1}{A_1A_2},\ \dots,\ v_k = \frac{S_{k-1}}{\prod\limits_{j = 1}^{k}A_j},
\label{V_k}
\ee
where $A_j \ne 0$, because $\beta_h > 0$. 

Ley us fix the number $n$. Then, from the necessary condition (\ref{NC}), which we can express as follows:
\be
\left(\alpha_h + \beta_h\right)^2 + \alpha_h\frac{\lambda_n^2}{g^2} = 0
\ee
one can obtain two roots for $\beta_h$:
\be
\beta_h = \beta_{h_{1,2}} = - \alpha_h \pm \frac{\lambda_n}{g}\sqrt{-\alpha_h}. 
\label{Beta_h}
\ee
Serie (\ref{H_SERIES}) is a polynomial of degree $N_h = n + 1$, iff $v_{N_h + 1} = 0$, i.e., $\left. S_{N_h} \right|_{\beta_h = \beta_{h_{1,2}}} = 0$, because $\left. C_{N_h + 2}\right|_{\beta_h = \beta_{h_{1,2}}} = 0$. Factorizing first $S_p$:
\bea
\left. S_{1} \right|_{\beta_h = \beta_{h_{1}}} &=& - \frac{\alpha_h (\lambda_n - \delta)}{64 g^4} \left(\lambda_n^2 - \delta^2 - 4g^2\right) \nonumber \\
&\times&\left(4g\sqrt{- \alpha_h} - \lambda_n\alpha_h + \delta\alpha_h\right), \\
\left. S_{2} \right|_{\beta_h = \beta_{h_{1}}} &=& - \frac{\alpha_h^2 (\lambda_n - \delta)}{2304 g^6} \left(\lambda_n^2 - \delta^2 - 4g^2\right)\left(\lambda_n^2 - \delta^2 - 8g^2\right) \nonumber \\
&\times&\left(6g\sqrt{- \alpha_h} - \lambda_n\alpha_h + \delta\alpha_h\right), \\
\left. S_{3} \right|_{\beta_h = \beta_{h_{1}}} &=& - \frac{\alpha_h^3 (\lambda_n - \delta)}{147456 g^8} \left(\lambda_n^2 - \delta^2 - 4g^2\right)\left(\lambda_n^2 - \delta^2 - 8g^2\right)\nonumber \\&\times&\left(\lambda_n^2 - \delta^2 - 12g^2\right) 
\left(8g\sqrt{- \alpha_h} - \lambda_n\alpha_h + \delta\alpha_h\right),
\label{S_p}
\eea
we suggest the form
\bea
\left. S_{p} \right|_{\beta_h = \beta_{h_{1}}} &=& - \frac{\alpha_h^p (\lambda_n - \delta)}{\left[(2g)^{p+1}(p+1)!\right]^2} \prod\limits_{j=0}^{p-1}\left(\lambda_n^2 - \lambda_j^2\right)\nonumber \\&\times&\left(2g(p + 1)\sqrt{- \alpha_h} - \lambda_n\alpha_h + \delta\alpha_h\right),
\label{S_p_Gen}
\eea
which is zero for $p = N_h$ due to factor $\lambda_n^2 - \lambda_j^2$ at $j = p - 1$. To demonstrate the suggested form by induction, we use (\ref{S_p}) for $p\mapsto p + 1$:
\be
S_{p + 1} = B_{p + 2} S_{p} + C_{p + 2} A_{p + 1} S_{p - 1}.
\ee  
Introducing $B_{p + 2}$, $C_{p + 2}$, $A_{p + 1}$ from (\ref{ABC1}), (\ref{ABC2}) to the right-hand side of the above relation, using (\ref{S_p_Gen}) for $S_p$ and $S_{p - 1}$ and evaluating at $\beta_h = \beta_{h_{1}}$, we obtain (\ref{S_p_Gen}) with $p\mapsto p + 1$. Thus, $\left. S_{N_h} \right|_{\beta_h = \beta_{h_{1}}} = 0$ for any fixed $n$ and we have a polynomial $h_{n + 1}(z)$ as in (\ref{h_n+1}). 

To show that the functions $u^{3,4}_n(\vec{r}, t)$ (\ref{uinv2-44})–(\ref{uinv2-4}) do not have a singularity at $r = 2|a|$, it suffices to prove that the following expression holds for a fixed number $n$:
\be
Y^\prime_n(2|a|) \operatorname{sign} a = \frac{\delta}{g\sqrt{2}}Y_n(2|a|).
\ee
For the sake of simplicity we consider $a>0$ and the upper sign in (\ref{Beta_h}). Then, the above relation is equivalent to 
\be
a\frac{\delta - \lambda_n}{g\sqrt{2}} \sum\limits_{k=0}^{n+1}v_k - \sum\limits_{k=0}^{n+1} k v_k = 0.
\label{ZERO}
\ee
Using (\ref{V_k}) and (\ref{S_p_Gen}), we obtain the form of $v_k$ in terms of gamma-functions:
\be
v_k = \frac{(-2a^2)^k}{a_\delta}\frac{a_\delta + k}{k!} \frac{\Gamma(n+2)}{\Gamma(n + 2 - k)} \frac{\Gamma(1 + \beta_h)}{\Gamma(1+ \beta_h + k)},
\ee
where we denote $a_\delta := a\frac{\delta + \lambda_n}{g\sqrt{2}}$. Then, using the properties of hypergeometric functions, one obtains:
\bea
\sum\limits_{k=0}^{n+1}v_k &=& {_2 F_2}\left(- n - 1,1 + a_\delta; 1 + \beta_h, a_\delta; 2a^2 \right), \\
\sum\limits_{k=0}^{n+1}k v_k &=& a\frac{1 + a_\delta}{g\sqrt{2}} \frac{\delta - \lambda_n}{1 + \beta_h} \nonumber\\
&&\times {_2 F_2}\left(- n, 2 + a_\delta; 2 + \beta_h, 1 + a_\delta; 2a^2 \right).
\eea
Factorization of (\ref{ZERO}) gives
\be
a\frac{\lambda_n^2 - \delta^2 - 4g^2(n + 1)}{g\sqrt{2}} \frac{{_1 F_1}\left(-n; 2+\beta_h; 2a^2\right)}{(1 + \beta_h)(\delta + \lambda_n)} = 0
\ee
for any natural $n$.

\bibliography{man}   

\end{document}